\PassOptionsToPackage{table}{xcolor}
\pdfoutput=1
\documentclass[acmsmall,screen,reviewa]{acmart}
\usepackage{tabularx} 
\usepackage{xspace}
\usepackage{xcolor}
\usepackage{amsmath}
\usepackage{algorithmic}
\usepackage{array}
\usepackage[caption=false,font=footnotesize]{subfig}
 \usepackage{grffile}
\usepackage{url}
\usepackage{lipsum}
\usepackage{textcomp}
\usepackage{tcolorbox}
\usepackage{multirow}
\usepackage{enumitem}
\usepackage{graphicx}

\usepackage{cleveref}

\Crefname{figure}{Figure}{Figures}
\crefname{figure}{Figure}{Figures}

\crefname{listing}{Listing}{Listings}
\Crefname{listing}{Listing}{Listings}

\Crefname{table}{Table}{Tables}
\crefname{table}{Table}{Tables}

\Crefname{box}{Figure}{Figures}
\crefname{box}{Figure}{Figures}

\crefformat{chapter}{\S~#2#1#3}
\crefmultiformat{chapter}{\S\S~#2#1#3}{ and~#2#1#3}{, #2#1#3}{, and~#2#1#3}

\crefformat{section}{Section~#2#1#3}
\crefmultiformat{section}{\S\S~#2#1#3}{ and~#2#1#3}{, #2#1#3}{, and~#2#1#3}

\usepackage{hyperref}

\usepackage{soul}

\newcommand{\ourTool}{\textit{Lingxi}\xspace}
\newcommand{\sweid}[1]{\texttt{#1}}

\newcommand{\prompt}[4]{
    \begin{figure}[htpb]
        \centering
        \begin{tcolorbox}[fontupper=\footnotesize,
            colframe=black, colback=white, coltitle=white, colbacktitle=black,
            title=#1 %, label=#3
        ]
            \textcolor{black}{
                #2
            }
        \end{tcolorbox}
        \caption{#4}
         \label{#3}
    \end{figure}
}

\begin{document}

% \title{Transferable Issue-Fixing Knowledge is All You Need: A Repo-Level Issue Resolution Framework Powered by Issue-Fixing Knowledge-Guided Scaling}
\title{Lingxi: Repository-Level Issue Resolution Framework Enhanced by Procedural Knowledge Guided Scaling}
\begin{abstract}

Driven by the advancements of Large Language Models (LLMs), LLM-powered agents are making significant improvements in software engineering tasks, yet struggle with complex, repository-level issue resolution. Existing agent-based methods have two key limitations. First, they lack of procedural knowledge (i.e., how an issue is fixed step-by-step and rationales behind it) to learn and leverage for issue resolution. Second, they rely on massive computational power to blindly explore the solution space.
To address those limitations, we propose \ourTool, an issue resolution framework that leverages procedural knowledge extracted from historical issue-fixing data to guide agents in solving repository-level issues. 
\ourTool first constructs this knowledge offline through a hierarchical abstraction mechanism, enabling agents to learn the how and why behind a fix, not just the final solution.
During online application, it employs a knowledge-driven scaling method that leverages the procedural knowledge of similar issues to intelligently analyze the target issue from multiple perspectives, in sharp contrast to undirected, brute-force exploration.
\ourTool successfully resolves 74.6\% of bugs on the SWE-bench Verified benchmark in Past@1 setting, outperforming five state-of-the-art techniques by a significant margin (5.4\% to 14.9\%).
Our comprehensive ablation study confirmed that the success of \ourTool comes directly from its use of procedural knowledge.
Without it, the performance gains from scaling alone is negligible.
Our qualitative study further shows that the ``design patterns $\&$ coding practices'' is the most critical knowledge aspect, and that the roles of different knowledge aspects switch across different stages (i.e., analysis, planning, and fixing).

\end{abstract}

% NEEDS fixing!

\author{Xu Yang}
\email{yangx4@myumanitoba.ca}
\affiliation{%
  \institution{University of Manitoba}
  \city{Winnipeg}
  \state{Manitoba}
  \country{Canada}
}

\author{Jiayuan Zhou}
\email{jiayuan.zhou1@huawei.com}
\affiliation{%
  \institution{Huawei}
  % \city{Winnipeg}
  % \state{Manitoba}
  \country{Canada}
}

\author{Michael Pacheco}
\email{michael.pacheco1@huawei.com}
\affiliation{%
  \institution{Huawei}
  \country{Canada}
}

\author{Wenhan Zhu}
\email{wenhan.zhu@huawei.com}
\affiliation{%
  \institution{Huawei}
  \country{Canada}
}

\author{Pengfei He}
\email{hep2@myumanitoba.ca}
\affiliation{%
  \institution{University of Manitoba}
  \city{Winnipeg}
  \state{Manitoba}
  \country{Canada}
}

\author{Shaowei Wang}
\authornote{Corresponding author.}
\email{shaowei.wang@umanitoba.ca}
\affiliation{%
  \institution{University of Manitoba}
  \city{Winnipeg}
  \state{Manitoba}
  \country{Canada}
}

\author{Kui Liu}
\email{kui.liu@huawei.com}
\affiliation{%
  \institution{Huawei Software Engineering Application Technology Lab}
  \city{Hangzhou}
  \country{China}
}

\author{Ruiqi Pan}
\email{panruiqi@huawei.com}
\affiliation{%
  \institution{Huawei Software Engineering Application Technology Lab}
  \city{Hangzhou}
  \country{China}
}



\maketitle
\section{Introduction}\label{sec:intro}

Driven by the advancement of Large Language Models (LLMs), a new generation of AI-powered agents is achieving substantial progress in software engineering, demonstrating significant capabilities in tasks like code completion~\cite{lu2022reacc} and generation~\cite{koziolek2024llm,parvez2021retrieval}.
However, a significant challenge remains: the automated resolution of repository-level issues. The complexity of this task stems from intricate module interactions and implicit business logic, which create an exponential solution space. To effectively address such issues, human developers draw upon deep procedural knowledge, encompassing both diagnostic and remediation strategies to achieve accurate problem understanding, guide strategic exploration, and ensure accurate patch implementation.%\wz{this reads human developers ensure accurate patch generation, alternative \textrightarrow implementation?}.

Current agent-based approaches attempt to mimic the developer debugging process by equipping agents with powerful backbone LLMs and an array of analytical tools to parse repository-specific context~\cite{wang2024openhands, moatless, yang2024swe}. However, a significant limitation of these methods is their focus on the state of the code when the issue is reported~\cite{chen2025swe,xia2024agentless}, often neglecting the significant value of historical issue-fixing data. Even when such historical data is considered, the focus typically remains on surface-level information (i.e., issue report-patch pair) rather than underlying key problem-solving procedural knowledge (e.g., root cause analysis and fix plan). This fundamental gap manifests in two key limitations of current approaches: 

%version 0907
% Prior studies have proposed techniques and frameworks to address repository-level issue resolution through carefully designed architectures, external tool integration, and repository knowledge construction~\cite{tang2025agent,wang2024openhands,gao2025trae}. When confronting the uncertainty inherent in complex problems, these methods typically employ \sw{should we say increasing the computational capability to explore solutions blindly (e.g., TTS)}diversity exploration strategies (such as test-time scaling~\cite{gao2025trae}\wz{should we add one more example}) to improve success rates, yet fail to capitalize on the value of development experience. Current methodologies exhibit significant limitations in the construction and application of experiential knowledge, manifesting primarily in two aspects:

% version 09-08 new angle
% From the perspective of knowledge construction, current methods for utilizing historical data are superficial. They treat historical issues as simple question-answer pairs: the raw text of the issue report and the final patch. This approach successfully captures the what (the problem and its solution) but completely omits the reusable how and why. It is akin to providing the math problem \(1+4*5+3 = 24\) without explaining the underlying procedure of operator precedence.
\noindent\textbf{Limitation 1: Lack of Transferable Procedural Knowledge from Historical Experience.}
While some prior works attempt to leverage historical data such as past issue reports~\cite{pabba2025semagent}, their utilization methods remain superficial. They treat historical issues as simple question-answer pairs: the raw text of the issue report and the final patch. This approach successfully captures the what (the problem and its solution) but omits the reusable how and why. It is akin to providing the math problem \(x^2+x-6=0, \) and answer \(x=2/-3\) without explaining the underlying procedure of factoring or the quadratic formula. This omission of the problem-solving method limits the ability to transfer this knowledge to solve other quadratic equations.
Similarly, issue reports and their patches fail to convey the crucial procedural knowledge applied by the developer, such as the diagnostic reasoning, exploration paths, and decision-making processes that led to the fix. The absence of such procedural knowledge is the primary barrier to effective knowledge transfer. As a result, while agents may retrieve textually ``similar cases'', they cannot learn the generalizable problem-solving strategies, leading to superficial imitation that fails in new scenarios (see~\Cref{motivation_scenario2} for more detail).

\noindent\textbf{Limitation 2: Knowledge-Blind Brute-Force Exploration.}
%Lack of Knowledge-Guided Exploration Processes
% For repository-level issues, complex module interactions and implicit business logic create an exponential solution space.
Without transferable procedural knowledge to provide direction, existing approaches default to a brute-force strategy, relying on massive computational power to blindly explore the solution space. This is typically done through indiscriminate sampling techniques like test-time scaling (TTS)~\cite{wang2024openhands, augment, antoniades2024swe,gao2025trae,test.time.scaling.huggingface,zhang2025and,test.time.scaling.nvidia}, which generates a diversity of patches by setting a high temperature or relying on different base models.
However, without guidance from repo-specific issue-fixing knowledge, this exploration is undirected and inefficient. Moreover, it is a ``many-to-one'' strategy that generates multiple candidates and hopes that one is correct (see~\Cref{motivation_scenario3} for more details). Consequently, this ad-hoc process is not only computationally expensive but also struggles to reliably converge on an effective solution for complex repository-level issues~\cite{gao2025trae}.

We propose \ourTool, an issue resolution framework that leverages procedural knowledge extracted from historical issue-fixing data to guide agents in solving repository-level issues. \ourTool is composed of three phases. First, it constructs transferable procedural knowledge from raw historical issue-fixing data (i.e., historical issue reports and theirs patches) through a hierarchical abstraction mechanism (addressing Limitation 1) offline. Second, during online application phase, it employs a novel knowledge-driven scaling method to analyze the issue from various perspectives guided by transferable procedural knowledge of similar issues, rather than relying on undirected sampling (addressing Limitation 2). Finally, the analysis report is sent for issue resolution. The key innovation lies in our knowledge-guided scaling approach, where multiple relevant historical issues with their associate transferable procedural knowledge are retrieved and analyzed in parallel to integrate complementary insights from different exploratory paths, where each might reveal a distinct facet of the issue.

We conducted comprehensive experiments on SWE-bench Verifed~\cite{swe.bench.verified}, which is a widely-used software issue resolution benchmark, containing 500 issue reports across 12 open source projects. Our results show that under the same foundation model setting, \ourTool achieves a Pass@1 resolution rate of 74.6\%, and significantly outperforms five state-of-the-art open scaffolding baselines (i.e., Openhands~\cite{wang2024openhands}, Augment Agent~\cite{augment}, SWE-search~\cite{moatless}, SWE-agent~\cite{yang2024swe} and its mini version mini-SWE-agent) by 5.4\% to 14.9\%.  We also conducted a comprehensive ablation study and observed that the use of transferable procedural knowledge is the backbone to boosts the performance of \ourTool, a significant improvement over using raw issue-patch pairs (68.3\%). The efficacy of scaling approach is critically dependent on knowledge guidance and only achieves significant performance improvement when guided by transferable knowledge.
We further conducted a qualitative study to understand to what extent does the procedural knowledge helps and observed that design patterns $\&$ coding practices are the most critical knowledge aspect. The relative importance of knowledge aspects clearly shifts across the analysis, planning, and fixing stages. Additionally, we also manually analyze the failure cases to provide insights for future works.

In summary, this paper makes the following contributions:
\vspace{-0.06in}
\begin{itemize}
    \item To the best of our knowledge, we are the first study to mine procedural knowledge and adopt it via a knowledge-guided mechanism for issue analysis and resolution.
    \item We conducted an extensive evaluation of our approach on a popular benchmark using three base models and compared it against five state-of-the-art baselines, demonstrating superior Pass@1 performance, with ablations quantifying each component’s contribution.
    \item We conducted a comprehensive qualitative study that identifies which knowledge types are most impactful at each stage (analysis, planning, fixing) and analyzes representative failure modes to inform future research.

\end{itemize}

\section{Background and Related Work}

%In this section, we briefly introduce the history of attempts to tackle the SWE-bench benchmark, focusing on the design shift and knowledge enhancements.

\subsection{Repository-level Issue Resolution with LLMs}
The early tools for repository-level issue resolution focused on implementing the complete end-to-end pipeline for automatic issue resolution.
Once complete pipelines were established, most approaches adopted a fixed-phase workflow consisting of reproduction, understanding, fixing, and review.
This workflow is commonly implemented using a ReAct-style agent~\cite{yao2023react}, where the LLM interacts with the environment in a loop with a fixed recursion limit. More recent work explores multi-agent systems that leverage repository artifacts to enhance performance~\cite{zhang2023survey}.

Initial research contributed significantly in completing the end-to-end pipeline while improving repository understanding and simplifying the testing process. To reduce the overhead of setting up the environment, subsequent work~\cite{wang2024openhands,antoniades2024swe} introduced pre-built Docker containers for each instance, which have since become a shared resource between tools.

More recent work explores test-time scaling (TTS), where multiple candidate patches are generated and evaluated, trading computation for performance.
A key challenge in TTS is to identify a correct patch among many candidates~\cite{gao2025trae}.
Trae~\cite{gao2025trae}, DARS~\cite{aggarwal2025dars}, and DEI~\cite{zhang2024diversity} address this with selector components, while Satori-SWE~\cite{zeng2025satori} fine-tunes an LLM to perform mutations during the TTS stage.
Another direction of research focuses on fine-tuning LLMs specifically for issue resolution.
SWE-smith~\cite{yang2025swe} expands SWE-style training data and trains specialized models.
SWE-Dev~\cite{wang2025swe} leverages test cases for fine-tuning, Lingma~\cite{ma2024lingma} extracts reasoning steps from issues such as training data, and SWE-search~\cite{antoniades2024swe} applies Monte Carlo tree search to improve agent trajectory selection.

In this work, we build on these studies and introduce a knowledge-guided scaling approach. Instead of directly applying TTS directly to patch generation, we apply it to the analysis phase, where transferable knowledge from historical issues is mined and guide the later patch generation phases.

\subsection{Leveraging Software Artifacts for Software Maintenance}

While source code is the core element of a software repository, research in mining software repositories has shown that many other artifacts can support software development~\cite{kalliamvakou2014promises}.
Examples include code change history (e.g., commits~\cite{hindle2008large}), development activities (e.g., issue trackers~\cite{bissyande2013got}), and external documentation~\cite{ding2014knowledge}.
These artifacts often capture information absent from the code itself, such as design decisions, trade-offs, and constraints~\cite{plosch2014value,rastkar2010summarizing}.
Incorporating such insights enables more comprehensive and practical solutions~\cite{bertram2010communication}.

In the context of repository-level issue resolution, prior work has explored improving the pipeline with additional software artifacts.
Notable examples include Agentless~\cite{xia2024agentless}, which uses pre-built repository-level graphs to improve repository understanding; AutoCodeRover~\cite{zhang2024autocoderover}, which provides the LLM with abstract syntax tree (AST) information;
SemAgent~\cite{pabba2025semagent}, which extracts related issue data for additional context;
and Code Graph Model~\cite{tao2025code}, which constructs code graphs to help LLMs better capture the issue context.

Unlike existing work, which focuses on the code-level context of the repository and the text-level context of past development artifacts,
\ourTool introduce a knowledge abstraction layer for mined knowledge based on historical issues.
This process allows us to provide structured and procedural information for issue resolution to guide the LLM in performing domain-specific tasks, complementing direct source code access and existing semantic extraction techniques.

\section{Motivated Example}\label{sec:challenges}

\begin{figure}
    \centering
    \includegraphics[width=1\linewidth]{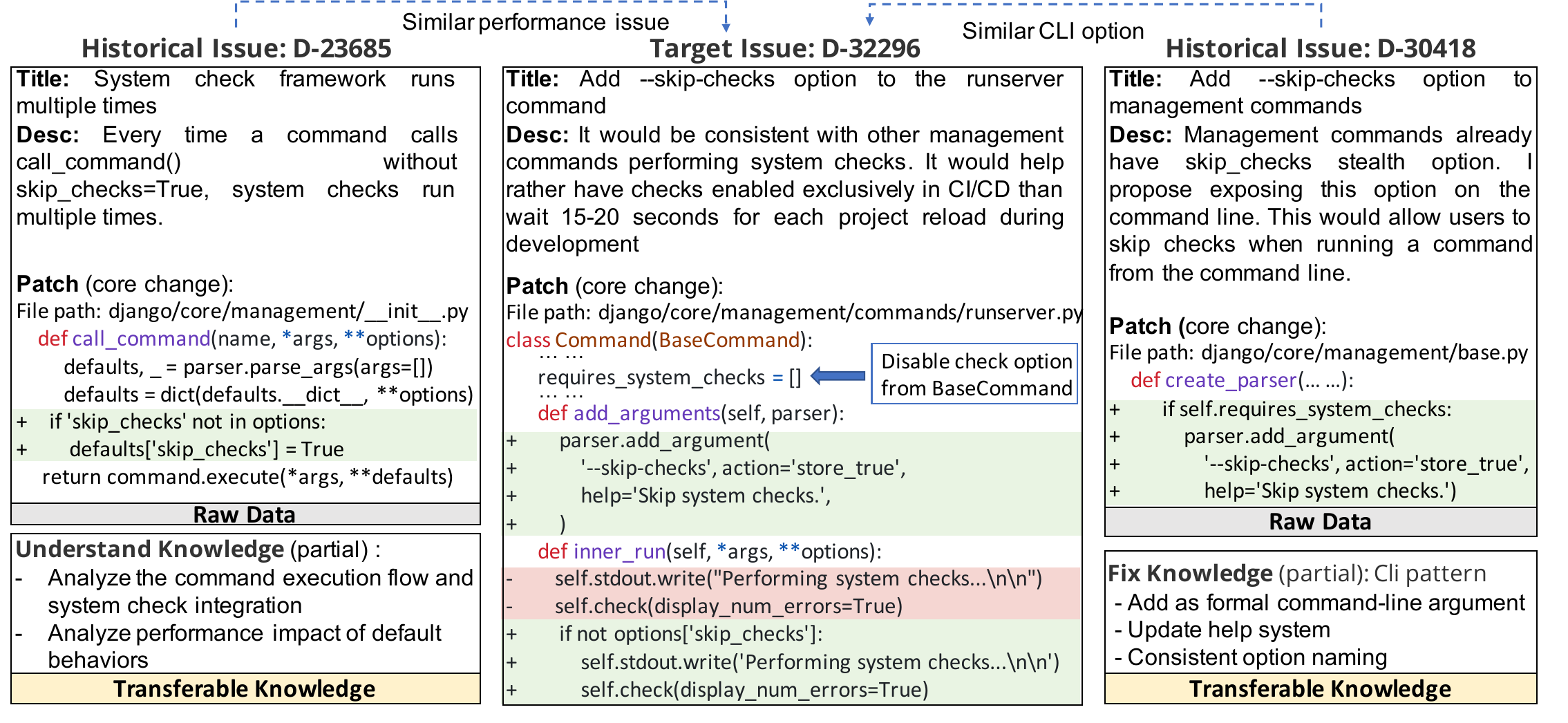}
    \caption{Target Django issue \sweid{D-32296} and its similar historical issues (\sweid{D-30418} and \sweid{D-23685})}
    \label{fig:motivation_example}
    \vspace{-0.3cm}
\end{figure}

% \begin{comment}
Repository-level issues require a comprehensive understanding that goes beyond surface functionality. Consider the Django issue \sweid{D-32296}\footnote{\url{https://code.djangoproject.com/ticket/32296}} (\Cref{fig:motivation_example}), which requests adding a \texttt{--skip-checks} option to the \texttt{runserver} command to bypass the time-consuming system check. We motivate our approach in the following three scenarios. 

\subsection{Scenario 1: The Naive Repository Exploration Approach}
\label{motivation_scenario1}

In the first scenario, an agent equipped with repository exploration tools might identify a straightforward solution: By examining the codebase structure, the agent observes that the parent \texttt{BaseCommand} class provides a standard mechanism through the \texttt{requires\_system\_checks} attribute. Following conventional patterns, the agent would attempt to set this attribute to \texttt{True} to enable the skip-checks functionality.
However, this approach would lead to a flawed implementation. A deeper inspection reveals that \texttt{runserver}  implements its own system checks through hardcoded calls within its \texttt{inner\_run()} method, bypassing the standard \texttt{BaseCommand} mechanism entirely.
The setting \texttt{requires\_system\_checks = []} is set to avoid potentially conflicting with this custom implementation.
The core challenge lies in recognizing and correctly handling this conflict between standard \texttt{BaseCommand} patterns and the specific local override implemented by \texttt{runserver}.

\begin{figure}
    \centering
    \includegraphics[width=1\linewidth]{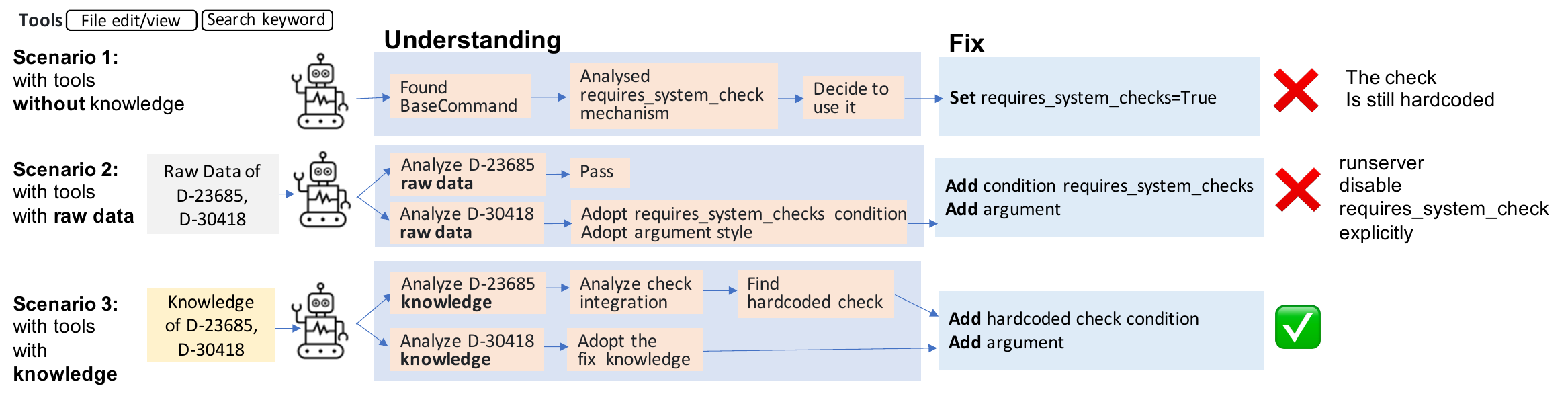}
    \caption{Example scenarios of an agent solving Django issue \sweid{D-32296} 1) with only tools, 2) with tools and raw data from historical issues, 3) with tools and transferable knowledge from historical issues}
    \label{fig:motivation_scenario}
    \vspace{-0.3cm}
\end{figure}

\subsection{Scenario 2: Learning from Raw Historical Issue-fixing Data}
\label{motivation_scenario2}

Recognizing the limitations of surface-level exploration, an agent might then turn to historical data for guidance, mimicking a developer's typical workflow of examining past similar issues. In this case, two relevant precedents emerge: \sweid{D-30418}, which implements the same CLI option to skip system checks, and \sweid{D-23685}, which also modifies system check behavior.
Yet directly applying the raw patches from these historical issues would fail to solve the current problem. The patch from \sweid{D-30418} proves ineffective because it relies on the standard mechanism that \texttt{runserver} deliberately disables, as identified in Scenario 1. Similarly, the patch from \sweid{D-23685} cannot be directly applied since it targets programmatic execution contexts rather than user CLI control.
The failure of these naive applications reveals a critical limitation: direct use of raw historical data without the underlying reasoning is often insufficient when the implementation context differs.%direct use of raw historical implementation data proves insufficient when the implementation context differs significantly from past cases.

\subsection{Scenario 3: Extracting and Fusing Transferable Knowledge}
\label{motivation_scenario3}

The key insight emerges when we move beyond copying raw data to extracting the underlying procedural knowledge from these historical cases. From \sweid{D-23685}, the essential value lies not in its specific patch, but in its problem analysis logic: ``analyzing command execution flow and performance impact of default behaviors''. When this analytical approach is applied to \texttt{runserver}, it reveals that hardcoded checks are indeed causing performance issues, suggesting that making these checks conditional would address the core problem.
To enable user control over this conditional behavior, the procedural knowledge from \sweid{D-30418} provides the implementation pattern for adding CLI arguments. While its specific patch was inapplicable, its approach to exposing check-skipping functionality through command-line options offers the necessary "how-to" knowledge.

To formalize this approach, let us define \textbf{transferable procedural knowledge}. First, \textbf{procedural knowledge} refers to the information that guides the entire issue resolution process, from initial understanding (e.g., root cause analysis) through to fix implementation (e.g., specific fix steps, test cases for validation). In other words, procedural knowledge encompasses the key information required to perform the end-to-end issue-fixing process. \textbf{Transferable procedural knowledge} refers to procedural knowledge that has been abstracted from issue-specific context and can be generalized to help resolve new issues. \textbf{We use the terms ``knowledge'' and ``procedural knowledge'' interchangeably throughout the following text.}

In addition, neither single piece of knowledge alone provides a complete solution as demonstrated above. The problem analysis from \sweid{D-23685} identifies the \textit{what} and \textit{why} of the problem (unconditional checks are problematic) but offers no direct implementation pattern for a CLI option. Conversely, \sweid{D-30418} provides the \textit{how} for the implementation but lacks the diagnostic insight needed to apply it to \texttt{runserver}'s unique context. Effective resolution therefore requires \textbf{fusing these complementary insights}: using the understanding knowledge from one issue to correctly identify the problem, and the fixing knowledge from another to correctly solve it. 
This example illustrates that agents must evolve beyond retrieving single, raw implementation-specific historical data and instead learn to abstract and fuse underlying reasoning chains from multiple sources to handle complex repository-level challenges effectively.

\section{Methodology}\label{sec:method}

\begin{figure}
    \centering
    \includegraphics[width=1\linewidth]{assets/workflow_v7.drawio.pdf}
    \caption{Workflow of \ourTool.}
    \label{fig:workflow}
    \vspace{-0.3cm}
\end{figure}

%\pf{\url{https://drive.google.com/file/d/1b1uQX_gTN7AJC4RCNdq4Rtg90hJglzIh/view?usp=sharing}}}
%现有的仓库级issue修复系统只依赖代码仓中的显性信息，例如代码，文件结构和测试文件。他们无法如开发者一般利用长期的维护代码仓积累下来的隐形开发经验来指导问题的修复。
%\sw{need to revisit after sub sectoins are done.}
%Current issue resolving approaches fail to abstract transferable procedural knowledge from raw historical issue-fixing records, which limits their ability to effectively reuse historical issue-fixing data to solve new issues, as demonstrated in \Cref{sec:callenges}. 

We propose \ourTool, an issue resolution framework that leverages procedural knowledge extracted from historical issue-fixing data to guide agents in solving repository-level issues. \Cref{fig:workflow} shows the workflow of \ourTool: 1) the knowledge is constructed offline, 2) when a new issue arrives, knowledge-guided analysis is performed, and 3) followed by the issue resolution process.

The framework establishes a novel knowledge-driven methodology that fundamentally differs from existing ad-hoc exploration approaches~\cite{antoniades2024swe, wang2024openhands}. Rather than relying on undirected sampling to explore the solution space, \ourTool extracts procedural knowledge from historical issue-fixing data, then uses this knowledge to guide both issue understanding and resolution processes. The key innovation lies in our knowledge-guided scaling approach, where multiple relevant historical issues are retrieved and analyzed in parallel to integrate complementary insights from different exploratory paths, where each might reveal a distinct facet of the issue.

\subsection{Transferable Procedural Knowledge Construction}\label{sec:knowledgeConstruction}
%\jy{Do not use evolution, hard to explain}\sw{Issue experience mining and abstraction? or Issue resolution knowledge construction}\jy{I propose to use development experiential knowledge mining. Then the raw data is historical development data, we first mine the development experience, then abstract it to generate development experiential knowledge. So data -> experience (for specific case) -> knowledge}\sw{agree}
% 知识agent的任务是对历史的issue-patch对进行知识提取并通用化。
Directly constructing transferable procedural knowledge is challenging. The procedural knowledge is buried deeply in issue report, its patch, and the repository. Without having a comprehensive understanding of the repo-specific information, it is impossible to mine accurate procedural knowledge. Therefore, we design a two-step process: 1) constructing issue-specific knowledge from historical issue-fixing pairs by navigating repository, and 2) generalizing it into general and reusable knowledge.

\subsubsection{Issue-Specific Knowledge Construction through Reverse Reasoning}
% 知识agent的第一个任务是建设历史issue的知识。这是因为历史issue的显性信息只有issue描述及其补丁，但是中间的推理过程/轨迹并不存在，而我们认为这些信息才是可以帮助后续issue解决的关键。因此，我们设计了知识agent，输入历史issue的issue描述和patch，并让其可以通过工具访问历史issue时间节点的仓库快照，让知识agent通过patch去进行反向推理，重建问题的解决路径。我们要求知识agent输出包括根因分析，

% -- by jy and claude
%Raw issue-patch pairs contain implicit experiential knowledge that requires systematic extraction to be useful. Simply providing more detailed issue descriptions does not guarantee better knowledge abstraction, as additional raw information may introduce noise rather than improve systematic understanding. 
%the two-step extraction help bridge the semantic gap between raw development data and the development knowledge.
%staged semantic transformation outperforms direct transformation when semantic distance is large.
%\sw{hmm, just come into my mind, why we call it ``reverse'', did we starting from the patch, or did we can both patch and issue report together.}\xy{we have both, but if we only have issue, agent might not able to reach the correct result, but when we have the result with the questions, agent can start from the result. We dont know if agent decide to start from issue or patch, but by providing the patch, agent can start from the patch}.
The first step involves mining issue-specific knowledge, which encompasses the detailed understanding and procedural information necessary to comprehend and resolve a particular issue. This includes knowledge for issue understanding (such as root cause analysis and behavioral comparison) and issue fixing (such as concrete fix steps and validation test cases), all derived from the corresponding issue report and patch.

This reverse mining approach is essential because historical issue-fixing data typically contain only the issue description (occasionally with logs and stack traces) and the final patch. The crucial procedural knowledge that bridges the gap between problem identification and solution implementation is rarely documented in existing repositories. Direct extraction of this procedural knowledge from issue reports without having a comprehensive understanding of repository information proves insufficient for a comprehensive understanding.

\begin{comment}
The first task is to mine issue-specific knowledge, i.e., the specific knowledge related to understanding (e.g., root cause analysis and behavioral comparison) and fixing an issue (e.g., concrete fix steps and test case for validation) for a particular issue, based on its issue report and patch. 
This is necessary because historical issues primarily only contain the initial issue description (sometimes having logs and stack traces) and the final patch, while the crucial intermediate information (i.e., procedural knowledge) related to issue understanding and fixing that connects them is merely documented. Mining this information directly from the issue report or repository is impossible. 
\end{comment}

To address this limitation, we design a knowledge agent that performs reverse reasoning from the available endpoints. Given an issue report and its corresponding patch, the agent reconstructs the implicit procedural knowledge that can guide the resolution process. The agent were equipped with the tools to explore and investigate the issue by navigating the code repository snapshot where the issue occurred. The agent will start from the patch to identify the modified files, browse those files, and reason why each of the file edits was related to the issues, and gradually construct the ground-truth issue fixing procedure, which mimics a developer's process for understanding issue-fixing. To ensure comprehensive knowledge extraction without omitting critical information, we instruct the agent to focus on 14 specific information types organized into two complementary categories: issue understanding and issue fixing.

\begin{comment}
To reconstruct this knowledge, we design a knowledge agent to perform reverse reasoning. Given an issue report and its final patch, the agent is designed to mine and reason the implicit knowledge that could help resolve the issue (see more details below). This process allows it to reconstruct the full knowledge required to resolve the issue as it records not only what changed, but why the change is necessary and how it was discovered. To ensure the agent collects comprehensive information and avoids missing key information, we design it to focus on 14 specific types of information, which are grouped into two families, i.e., issue understanding and issue fixing.
\end{comment}

\begin{itemize}

    \item \textbf{\underline{Issue understanding:}} we collect 8 types of information that are critical for problem comprehension and localization~\cite{bettenburg2008makes,wang2014version,chaparro2017detecting,chen2021pathidea}, which includes 1) \emph{Relevant architecture} that helps locate the issue within the system architecture; 2) \emph{Dependency analysis} of imports, inheritance, and call-site relations to help identify potential affected components; 3) \emph{Call trace} helps reconstruct the execution/order of relevant subsystems, files, and methods; 4) \emph{Bug categorization} (e.g., off-by-one, misuse of API) helps with precise localization; 5) \emph{Example usage context} explains when/why the affected component is used in practice; 6) \emph{Concrete root cause analysis} provides detailed specific conditions and logic that trigger failure; 7) \emph{Current vs.\ expected behavior} contrasts incorrect behavior with intent and helps to specify overlooked nuances; 8) \emph{Feature or functionality} helps identify relevant components that might require corresponding updates or testing.
    
    \item \textbf{\underline{Issue fixing:}} encompasses 6 types of information that is critical for fixing issues~\cite{yin2011fixes,ko2006exploratory,saff2004experimental}: 1) \emph{Fix pattern} explains where and why the patch works; 2) \emph{Concrete fix steps} explain how to fix an issue in concrete steps; 3) \emph{Fix checklist} explains necessary steps to verify during debugging and implementation; 4) \emph{Test case} is used to guide bug reproduction and validate the repair; 5) \emph{Resolution conclusion} summarizes the unexpected behavior of the issue, its scope, fix location, and applied resolution pattern; 6) \emph{Design patterns \& code practices} help ensure the solution adheres to established architectural principles and facilitates seamless integration with existing system components.
    
%1) concrete fix steps that provide numbered, 2) actionable instructions for applying the solution, and 3) test cases that validate the repair's effectiveness. \xy{should we simply here?} \sw{this is one of the core novelty. we should add rationales on why we choose them, and add citations to support our decision.}
\end{itemize}

%\sw{check if this is correct or not.}\sw{any specific prompt engineering or agent design principle we used? if we have, mention it}\sw{if we have space, put a prompt to showcase to make it more interesting and concrete.}
%Note that reverse reasoning of the 14 aspects of information is challenging, since almost all information are not directly documented, but buried deeply in the repository. Directly prompting LLMs to infer them is not practical without sufficient context information from the repository. Therefore, to achieve this, we equip the knowledge agent with necessary tools to investigate the repository. More specifically, the agent is designed to perform the following process to mimic developers' issue fixing process: 1) traces and investigates execution paths to identify failure points, 2) identifies root causes within their architectural context, and documents the concrete repair implementation steps. During this process, the agent actively collect necessary information to generate twelve aspects of issue-specific information.

\subsubsection{Transferable Knowledge Construction through Abstraction}
%知识agent的第二个任务是将具体的修复抽象为可重用的模式。这是因为在第一阶段中我们得到的issue specific经验与那个历史issue强相关，往往只能够被用于帮助解决特定的issue，而无法转移和重新利用，因此我们需要将其转化为更加通用的development knowledge. 实证软件工程研究已确定了一些特定的信息类别，它们能够显著提高开发人员解决问题的效率。我们遵循这些原则，从每个issue experience中抽象出最关键的知识要素。

The knowledge agent's second task is to abstract issue-specific knowledge into generalizable and transferable one. This transformation is essential because the issue-specific experience remains strongly coupled to a specific issue, limiting applicability to similar but non-identical problems. Therefore, we prompt the knowledge agent to abstract each type of issue-specific information into more general descriptions.  \Cref{fig:prompt_abstraction} presents the prompt template. The abstraction process targets two key objectives: 1) extracting generalizable patterns from specific problem contexts, and 2) ensuring the resulting knowledge remains actionable and educational for developers unfamiliar with the original system context. We also have requirements on structure and organization (e.g., using XML format) so that the knowledge can be used by following process seamlessly. During abstraction, we keep all aspects from issue-specific knowledge.
In \Cref{tab:knowledge_exp}, we show an example constructed knowledge for \sweid{sympy-15872}\footnote{\url{https://github.com/sympy/sympy/issues/15872}}.

\begin{table}[htbp]
\caption{Generated transferable knowledge for \sweid{sympy-15872}, some aspects are omitted due to space constraints. For complete information please refer to our replication package.}
\centering
\renewcommand{\arraystretch}{1.1} % overall row height
\newcommand{\fakeitem}{\textbullet\hspace{0.5em}}
\label{tab:knowledge_exp}
\resizebox{\textwidth}{!}{%
\begin{tabular}{@{}p{3.5cm}p{14cm}@{}}
\toprule
\textbf{Aspect} & \textbf{Details} \\
\midrule
General Root \newline Cause Analysis Steps & 
    \fakeitem \textbf{Identify the problematic behavior:} Observe inconsistent rank calculations for powered matrices; compare results with and without simplification.\newline
    \fakeitem \textbf{Isolate the conditions:} Create test matrices; apply matrix operations (powers); compare rank calculations.\newline
    \fakeitem \textbf{Analyze the computation flow:} Matrix construction and operations; rank calculation process; expression simplification handling.\newline
    \fakeitem \textbf{Verify correct behavior:} Create test cases with known correct results; compare against mathematical expectations; document correct behavior patterns.
    \\
\midrule
Bug \newline Categorization & 
    \fakeitem Type: Mathematical Computation Bug \newline
    \fakeitem Subtype: Expression Simplification Issue \newline
    \fakeitem Category: Matrix Operation Correctness \newline
    \fakeitem Impact Level: Core Functionality \newline
    \fakeitem Scope: Matrix Rank Calculation System
    \\
\midrule
Relevant\newline Architecture &
    \fakeitem \textbf{Hierarchy:} SymPy Core Math Library $\rightarrow$ Matrix System Layer $\rightarrow$ Matrix Operations Module $\rightarrow$ Rank Calculation Subsystem $\rightarrow$ (Matrix Power Operations, Expression Simplification System) \newline
    \fakeitem \textbf{Key Patterns:} Object-oriented matrix representation; mathematical expression tree structure; computation result caching system; expression simplification pipeline
    \\
\midrule
Dependency \newline Analysis &
    \fakeitem Matrix Base Component: core matrix operations, matrix construction \newline
    \fakeitem Matrix Operations Handler: matrix arithmetic, power operations \newline
    \fakeitem Rank Calculation System: computes matrix ranks, handles simplification \newline
    \fakeitem Test Framework: validates operations, ensures correctness
    \\
\midrule
Specific Classes/\newline Functions/Methods &
    \fakeitem \textbf{Classes:} Matrix, MatrixBase \newline
    \fakeitem \textbf{Methods:} rank(), Matrix.eye(), matrix power operator (**) \newline
    \fakeitem \textbf{Functions:} \texttt{test\_issue\_15872()}, matrix construction methods, simplification routines
    \\
\midrule
Feature / Functionality\newline of Issue &
    \fakeitem \textbf{Core:} Determines linear independence; calculates matrix ranks; handles matrix power operations \newline
    \fakeitem \textbf{Use Cases:} Linear equation solving; linear independence verification; matrix property analysis; linear transformation studies
    \\
\midrule
General\newline Fix Pattern &
    \fakeitem \textbf{Test-Driven Fix Approach:} Add comprehensive tests; verify behavior; document expected results \newline
    \fakeitem \textbf{Validation Pattern:} Test original matrix properties; verify powered matrix behavior; ensure consistent results
    \\
\midrule
General \newline Fix Checklist &
    \fakeitem Verify matrix initialization correctness \newline
    \fakeitem Check basic matrix operations \newline
    \fakeitem Validate rank calculations \newline
    \fakeitem Test power operations \newline
    \fakeitem Verify simplification behavior \newline
    \fakeitem Ensure consistent results \newline
    \fakeitem Document expected behavior \newline
    \fakeitem Add regression tests
    \\
\midrule
Design Patterns \&\newline Code Practices &
    \fakeitem \textbf{Test-First Development:} Write failing tests first; implement fixes to pass tests; verify edge cases \newline
    \fakeitem \textbf{Matrix Operation Patterns:} Proper construction; consistent operation handling; thorough validation \newline
    \fakeitem \textbf{Expression Handling:} Proper simplification; consistent evaluation; clear computation flow
    \\
% \midrule
% Additional\newline Concepts &
%     \fakeitem Linear Algebra: matrix rank properties, linear independence, matrix power behavior \newline
%     \fakeitem Symbolic Computation: expression simplification, term reduction, symbolic manipulation \newline
%     \fakeitem Testing Best Practices: edge case coverage, mathematical correctness, regression prevention \newline
%     \fakeitem System Architecture: matrix representation, operation workflow, computation pipeline
%     \\
\bottomrule
\end{tabular}%
}

\end{table}

%\sw{ not clear how this is been done? need improve. I would show the prompt, and briefly explain the prompt and its rationale, if it is too long, we can only keep the key component}

\prompt{Prompt Template: Knowledge Abstraction}
{%
  \underline{\textbf{\#Instruction}} \\
  Given the following issue analysis report of a historic issue, generate a comprehensive summary based on the entire report - with key knowledge and background information that is necessary or helpful in solving this issue. Your goal is to abstract the information from the original issue analysis report to a more general description. Imagine this is for a newcomer to the system, and try to summarize the key knowledge that the newcomer needs to know so that they can solve this issue, and also apply the knowledge to similar issues in the future. Your generated summary must follow the following aspects:\\
  1) General Root Cause Analysis Steps \\
  $\dots$\\
  Here is the issue analysis:\\ <historic\_analysis> \{$original\_issue\_report$\}+\{$issue\_specific\_knowledge$\}</historic\_analysis>\\
  Here is the patch info: \\
  <historic\_patch> \textbf{\{$patch$\}}  </historic\_patch> \\
  %\tcblower
  \underline{\textbf{\#Suffix}} \\
  Verify your generated analysis follows the characteristics below:\\
     - \textbf{General and abstracted}: Abstract information from the original issue analysis, transforming it into a more general description focusing on the core concepts and principles rather than specific details. The goal is to transform the original analysis into a broad overview, that can be applied to a wider range of similar issues.\\
     - \textbf{Actionable, Educational and Transferable}: Frame your insights in a way that is useful and easily understandable to future developers and newcomers to the system, as they will use your analysis to understand this issue, and apply its knowledge for fixing future similar issues. \\
     - \textbf{Structured and organized}: $\dots$ \\
     - \textbf{Insightful}: $\dots$
 }{fig:prompt_abstraction}{The prompt template for knowledge abstraction. Note that for saving space, we only keep the key component and use $\dots$ to denote the saving sentences.}

%\sw{the following description does not helps, remove it. }
%Root cause analysis patterns become systematic debugging procedures applicable to similar issue categories. Architectural relationships and component interactions reveal system-level design principles that guide future problem-solving. Concrete fix implementations abstract into generalized repair strategies with clear applicability conditions. Domain-specific constraints and conventions provide the contextual understanding necessary for correct fixes across similar scenarios. \sw{so the 12 aspect abstracted into 12 more general principles?}Through this abstraction process, specific instances transform into broader principles. \sw{what is this?} For example, adding a null check at line 234 evolves into a general principle about defensive programming in authentication pipelines. 

%\xy{fill all aspects for both steps}

%%We randomly sample x\% of generated knowledge entries across different repositories and knowledge dimensions to ensure representative coverage. Three senior software engineers (with 8+ years experience) independently evaluate each sample using predefined criteria: (1) accuracy of extracted information, (2) appropriateness of abstraction level, (3) transferability across contexts, and (4) completeness of knowledge dimensions.

% \sw{revisit the title}\xy{lgtm}
\subsection{Knowledge-guided Issue Analysis Scaling}
%\jy{suggest to move it to knowledge constcution. Currently there is no idea which two are the Dual-phase. At least need to clearly say phase 1: xxx, which extract k similar knowledge, and then do the analysis seperately. THen Phase 2: xxx, Infuse the analysis result.} \sw{I would skip the dual-phase wording.}

%\sw{need to polish after all subsections are done}
Existing approaches often compensate for a lack of insight by increasing computational power to brute-force explore the solution space, typically through blind sampling techniques like test-time scaling (TTS)~\cite{wang2024openhands, augment, antoniades2024swe,gao2025trae}. To address this inefficiency, we introduce a novel knowledge-guided mechanism for the issue analysis stage. 
Our approach first retrieves multiple relevant historical issues and their associated transferable knowledge (\textbf{Context-Aware Knowledge Retrieval}). 
%\jy{Review}\sw{good}
Each unit of knowledge is then assigned to a dedicated analysis agent, allowing the issue analysis to scale up effectively through multiple parallel instances (\textbf{Parallel Knowledge-guided Analysis}), meanwhile ensuring its exploration efficiency through knowledge-guided investigation. 
This multi-perspective exploration enables \ourTool to comprehend the target issue from diverse perspectives, as each knowledge source suggests a different solution path. Finally, the individual analysis reports undergo a comprehensive integration and refinement process that systematically consolidates valuable insights while eliminating noise (\textbf{Analysis Synthesis}).%Finally, the individual analysis reports are infused and refined by filtering out irrelevant information.

%\subsubsection{Issue Summarization}\label{sec:issuesummarization}
%\sw{is this done knowledge construction stage or offline. if so, merge it into knowledge construction stage. knowledge application stage should be only online process.}

\subsubsection{Context-Aware Knowledge Retrieval}
%\sw{merge this issue summarization into retrieval subsection}
Raw issue reports often contain significant noise, such as verbose templates, extraneous system configurations, multi-page stack traces, and extended discussion threads. This noise impedes the effective identification of relevant historical issues. To address this, we employ a structured summarization technique using LLMs to extract four critical elements~\cite{bettenburg2008makes}: 1) the core problem, 2) key technical details and error messages, 3) expected versus actual behavior, and 4) relevant code components. These summaries are then processed by an embedding model and stored in the issue-fixing knowledge database to facilitate efficient retrieval.

When a new issue is submitted, we first generate its structured summary using the same method. This summary is encoded into a high-dimensional embedding vector, which serves as a query to retrieve the top-$N$ most similar historical issues from the database. A re-ranking model subsequently performs a fine-grained relevancy assessment to re-order the initial results~\cite{gao2024llm}. The final output is a curated set of the $N$ most relevant historical issues along with their associated knowledge, providing diverse yet focused contextual perspectives for analysis.

\subsubsection{Parallel Knowledge-guided Analysis}

A straightforward approach would be to dump all the retrieved issue reports along with their bug-fixing knowledge into a single long prompt for the agent to analyze. However, this is problematic because LLMs often fail to use long contexts effectively and may miss crucial information buried within them. This ``Lost in the Middle'' problem can significantly hinder the effectiveness of LLMs~\cite{liu2023lost,li2024long}).

To overcome this limitation, we employ multiple analysis agents, where each agent is responsible for analyzing the target issue guided by one relevant issue knowledge from a distinct perspective. Specifically, each agent receives the target issue along with one relevant historical issue and its corresponding knowledge and produces an analysis report. By providing this focused context, each agent generates a detailed analysis report, ensuring all key information is fully considered. This parallel approach prevents the model from being overwhelmed by a single, massive prompt and allows for a more comprehensive, multi-faceted analysis and can be scaled up easily.  

%\sw{is this put in wrong place? did we use }We employ ensemble analysis where multiple problem decoder agents examine the issue through different knowledge-informed lenses. Each decoder receives the current issue along with different historical knowledge that serves as its analytical framework. This knowledge serve as the reference how the decoder approaches the problem.

\subsubsection{Analysis Synthesis}
%\jy{Do we call it aggregated synthesis or knowledge infusion? need to pick a word.}
%\sw{criteria: }
We perform the analysis synthesis on all analysis reports to produce a single consolidated report by leveraging an LLM, which is instructed to aggregate the the reports from the following four criteria:
% \sw{xu: can you check on this, i assume the examples you list here are only subset, not the complete set. so i use term ``such as''}\xy{correct, just to explain what does it means :)}
\begin{itemize}
    \item \textbf{Supportive} — Keep only the analysis that provides \emph{actionable evidence}, such as a reproducible trigger/log, precise code pointers (file:line/symbol/call path), and a hypothesis–evidence link.
    \item \textbf{Complementarity} — Keep only the non-overlapping insights that introduce new failure modes or affected components.
    \item \textbf{Accuracy} — Keep claims validated by the current code/tests/traces; prefer line-anchored evidence; drop contradictions or speculation.
    \item \textbf{Completeness} — Ensure the synthesis report covers key elements, such as root cause, scope/impact, and fix preconditions/constraints.
\end{itemize}

%\textbf{4. Completeness} (gaps no single decoder addressed). 
% This synthesis critically evaluates evidence strength, resolves competing hypotheses, and combines complementary insights into a coherent narrative. 

These criteria ensure that the aggregation process produces a single, consolidated analysis from multiple knowledge sources. The aggregation addresses the issue where individual analyses might miss key information due to limited perspectives, while filtering out irrelevant details. After this step, only the most relevant information is retained and carried forward to the following phases.

\subsection{Issue Resolution}
% \sw{better to consistent with agent actions in \Cref{fig:workflow}}\xy{synced}
% \sw{Xu: is this part of agent? please revise}\xy{yes, revised}
For issue resolution, we follow the process adopted in previous approaches~\cite{wang2024openhands, antoniades2024swe} and develop two agents: 1) \textbf{Planning agent}, which transforms the issue analysis report into a repair plan through exploration, reproduction, and fix planning. This agent produces a verified root cause, a line-level code-change plan, and reasoning that links the changes to the identified problem; and 2) \textbf{Fixing agent}, which executes the fix plan with systematic verification through code editing, reproduction testing, and iterative refinement, outputting the applied patches and test results.

\subsection{Principles of Agent Design}

We introduce the principles of agent design involved in \ourTool in this section. 

\subsubsection{General agent architecture}
All our agents follow the ReAct agent~\cite{yao2023react} design by following previous work~\cite{aggarwal2025dars,antoniades2024swe,gao2025trae,ge2023openagi,augment,xia2024agentless,wang2024openhands,yang2024swe}. Except for that, all agents were prompted to follow a structured reasoning output after each tool calling, including observation, considering alternative actions, and reasoning next actions. This reasoning output forces explicit processing rather than blind tool call chaining and creates interpretable reasoning traces for debugging.

% \subsubsection{Specific agent}\sw{we can merge this into above phases if we need space.}
% Our framework separates the repair process into three specialized agents, each optimizing for its specific task without conflating concerns.
% \begin{itemize}
%     \item Problem Decoder (Ensemble with Top-3 Knowledge). Analyzes issues to identify root causes using diverse historical knowledge perspectives. Three decoders examine the issue through different knowledge lenses, producing analyses synthesized into unified understanding. Outputs problem statement, root cause, current behavior analysis, and expected behavior specification.\sw{we don't have this problem decoder in workflow, is it replaced by analyze agent?}\pf{yes,decoder->analysis, mapper->plan, solver->fix}
%     \item Solution Mapper. Solution Mapper transforms problem understanding into repair plans through exploration, reproduction, and fix planning. Produces verified root cause, line-level change specifications, and reasoning linking changes to the problem.
%     \item Problem Solver. Problem Solver executes repair plans with systematic verification through implementation, testing, and iterative refinement. Outputs applied patches and test results.
% \end{itemize}

\subsubsection{Agent Tools}\label{sec:toolused}
\sloppy Similar to previous work, we employ our agents with the following tools: \texttt{view\_directory} for adaptive repository traversal, \texttt{view\_file\_content} for code repository file browsing, \texttt{search\_files\_by\_keywords} for semantic code search via \texttt{ripgrep}~\cite{ripgrep}, and \texttt{str\_replace\_editor} for code editing.

\subsubsection{Context Management}
We compress information passing through multi-agent pipelines to prevent information overload. More specifically, we extract observations and reasoning while discarding verbose tool calling outputs. This preserves essential findings while removing redundant content that causes attention dilution when context windows become too large. We summarize the flowing information between agents rather than complete conversation histories ~\cite{xia2024agentless,wang2024openhands,yang2024swe}.%

\section{Experimental Settings}
\subsection{Research Questions}
Our evaluation addresses three research questions:
\begin{itemize}
    \item \textbf{RQ1}: How does \ourTool perform compared to state-of-the-art repository-level issue resolution techniques?
    \item \textbf{RQ2}: What is the individual contribution of each major component to \ourTool?
    \item \textbf{RQ3}: To what extent does transferable knowledge help improve issue resolution performance?
\end{itemize}

\subsection{Benchmark and Dataset}
We evaluate \ourTool on SWE-bench Verified, a curated subset of 500 GitHub issues selected from the original SWE-bench dataset~\cite{jimenez2023swe} of 2,294 real-world issues. Professional developers manually verified each issue in this subset to ensure unambiguous problem descriptions, appropriate scope, and reliable test coverage~\cite{swe.bench.verified}.
Each issue in the benchmark consists of an issue description in natural language and the original repository snapshot at the issue creation time. The benchmark provides golden tests that verify whether a generated patch correctly resolves the issue. All evaluated techniques receive identical inputs: the issue description and the repository codebase, ensuring fair comparison.
To prevent temporal leakage, the historical issue candidate pool is limited to issues created before the target issue’s creation time.
% \jy{Need explicit introduction on how we build the experience data, to avoid potential data leakage concerns from reviewers.}\jy{Also mention: for each issue, SWE-bench Verified provide docker enviornment where the codebase is the snapshot before the issue is created}\xy{@jy, revised}

\subsection{Evaluation Metrics}
We adopt Pass@1 as our evaluation metric to evaluate the effectiveness of the Lingxi and other techniques by following previous studies~\cite{wang2024openhands, augment, antoniades2024swe,gao2025trae}. Pass@1 measures the percentage of issues successfully resolved by each technique's generated patch and does not attempt the same task issue more than once. An issue is considered resolved if the generated patch passes all associated golden tests in SWE-bench. This metric represents the primary evaluation criterion, as it directly measures the practical utility of automated repair systems. A higher Pass@1 score indicates better effectiveness.

\subsection{Base Models and SOTA Baselines}
In our experiments, we evaluate \ourTool on three recent high-performing LLMs (2 commercial models, and 1 openweighted model).\footnote{Note that the experiment with commercial models are completed before Sept., 2025} To establish state-of-the-art baselines, we refer to the SWE-bench leaderboard,\footnote{\url{https://www.swebench.com/}} which ranks repository-level issue resolution techniques. To ensure a fair comparison focused on methodology rather than model capabilities, we only select top-performing techniques that utilize the same base models as our study. We explicitly exclude methods that combine multiple model families to isolate the impact of our approach. This selection process yields the following baselines:

%\wz{maybe make it a table?}
\begin{itemize}
    \item \textbf{OpenHands} \cite{wang2024openhands}: is an agent-based framework with planning mechanisms and tool integration, representing the standard agent orchestration approach. 
    \item \textbf{Augment Code (Augment)}~\cite{augment}: applies a prompting-driven ensemble strategy modeled using the LLM-as-a-judge approach. It guides the LLM to compare candidate patches with the issue description and determine the most appropriate one.
    \item \textbf{SWE-agent/mini-SWE-agent:} \cite{yang2024swe}: SWE-agent introduces a custom agent–computer interface that enables language model agents to autonomously navigate codebases, edit files, and run tests, significantly improving their performance on software engineering tasks.
    mini-SWE-agent is a 100 line code version introduced by the same team of SWE-agent, and focuses on a minimal, bare-bones architecture. 
    \item \textbf{SWE-search }\cite{antoniades2024swe}: is a multi-agent framework that combines Monte Carlo Tree Search with self-improvement mechanisms, enabling LLM-based software agents to iteratively explore, evaluate, and refine strategies for repository-level software engineering tasks.
\end{itemize}

\subsection{Implementation Detail}

\ourTool interfaces with each LLM through its official API. For knowledge retrieval, we embed the target issue summary with Qwen3-Embedding-8B~\cite{zhang2025qwen3} and retrieve the top-20 nearest historical issues by cosine similarity; we then rerank these candidates with Qwen3-Reranker-4B~\cite{zhang2025qwen3} and keep the top~3. 
%To prevent temporal leakage, the historical issue candidate pool is limited to issues created before the target issue’s creation time.

\section{Results}
\subsection{RQ1: Overall Effectiveness of \ourTool}
\subsubsection{Approach}
To answer RQ1, we compare the performance of \ourTool with four state-of-the-art SWE-bench baselines (i.e., OpenHands, Augment, SWE-Agent, SWE-Search).

\subsubsection{Results}

\textbf{\ourTool contently outperforms all studied SOTA baselines across all base models. For instance, \ourTool achieves a Pass@1 of 74.6\%, yielding performance improvements ranging from 5.4\% to 14.9\% over baselines.} \Cref{tab:resolved_rates} presents the comparison between \ourTool and baselines when integrated with different base models. As observed, \ourTool consistently outperforms all SOTA baselines. For example, for the most outstanding model, \ourTool achieves a Pass@1 of 74.60\%, while the Pass@1 of baselines ranges from 66.6\% to 70.4\%. We observe similar patterns for GPT-4.1 mini and Kimi-K2, which indicates the generalizability of \ourTool across different base models. Note that \ourTool achieves its most significant improvements when applied to weaker base models. For instance, on GPT-4.1 mini, \ourTool increases the Pass@1 rate from 23.94\% (mini-SWE-agent) to 45.20\%. In contrast, the improvement on the best model is more modest, rising from 64.93\% to 74.60\%.

\begin{table}[t]
  \centering
  \caption{Effectiveness comparison between \ourTool and baselines across different base models.}
    \centering
    \begin{tabular}{@{}l l r@{}}
    \toprule
    \textbf{Approach} & \textbf{Base Model} & \textbf{Pass@1(\%)} \\
    \midrule
    mini-SWE-agent         & GPT-4.1 mini & 23.94 \\
    \ourTool        & GPT-4.1 mini & \textbf{45.20} \\
    \midrule
    mini-SWE-agent        & Kimi-K2 & 43.80 \\
    OpenHands         & Kimi-K2 & 65.40 \\
    \ourTool        & Kimi-K2 & \textbf{70.33} \\
    \midrule
    mini-SWE-agent         & Claude 4 Sonnet & 64.93\\
    SWE-agent         & Claude 4 Sonnet & 66.60 \\
    SWE-search        & Claude 4 Sonnet & 70.08 \\
    OpenHands         & Claude 4 Sonnet & 70.40 \\
    Augment Agent v1  & Claude 4 Sonnet & 70.40 \\
    \ourTool            & Claude 4 Sonnet & \textbf{74.60} \\
    \bottomrule
    \end{tabular}

    \label{tab:resolved_rates}
\end{table}

\newpage
\subsection{RQ2: Ablation Study}

\subsubsection{Approach}

\begin{table}[t]
\centering
\caption{Ablation study results on SWE-bench-Verified using the best performance model in RQ1.}
\begin{tabular}{@{}lp{3in}cc@{}}
\toprule
\textbf{Model}    & \textbf{Setting}               & \textbf{Pass@1 (\%)} & \textbf{$\Delta$ (\%)}                    \\ \midrule
\ourTool       & Full version                              & \textbf{74.80}       & \textcolor{gray}{--}   \\ \midrule
$\ourTool_{woS}$  & without analysis scaling            & 72.48                & \textcolor{red}{-3.10} \\
$\ourTool_{woT}$ & without transferable procedural knowledge, instead using raw issue-patch pairs & 68.30                & \textcolor{red}{-8.69} \\
$\ourTool_{woK}$  & without any knowledge              & 67.74                & \textcolor{red}{-9.44} \\
$\ourTool_{woSK}$ & without any knowledge and analysis scaling                      & 67.43                & \textcolor{red}{-9.87} \\ \bottomrule
\end{tabular}
\label{tab:ablation}
\end{table}
To precisely quantify the contribution of each component in \ourTool, we conduct a comprehensive ablation study. We systematically remove or replace key elements to measure their impact on performance. Due to the high cost of evaluating on all 500 issues, we instead use a statistically rigorous sample. We randomly selected 218 instances, a sample size that provides a 95\% confidence level with a 5\% margin of error, ensuring our findings are generalizable to the entire SWE-bench Verified dataset. Based on the results from RQ1, we choose the model that achieve the best performance with \ourTool to conduct the ablation study.

Our ablation study focuses on two critical design elements:
\begin{itemize}
\item \textbf{Transferable knowledge:} We assess the effectiveness of different types of knowledge by comparing the following two variants with \ourTool.
$\ourTool_{\text{woT}}$, where we replace the transferable knowledge with raw issue-patch pairs, while keeping all other components identical. 
$\ourTool_{\text{woK}}$, where we remove knowledge retrieval entirely. The target issue report is sent directly to the analysis agent, though the analysis scaling component is retained. In other words, we scale up the issue analysis on the same issue report without any knowledge guided.

\item  \textbf{Knowledge-guided analysis scaling:} To measure the impact of scaling itself, we compare \ourTool with $\ourTool_{\text{woS}}$, a variant that uses only the top-1 relevant issue's knowledge and disables analysis scaling. 
To measure the synergy between knowledge and scaling, we compare $\ourTool_{\text{woK}}$ (no knowledge, scaling enabled) with $\ourTool_{\text{woSK}}$ (no knowledge, scaling disabled), which isolates the effect of scaling in the absence of knowledge guidance.

\end{itemize}

\subsubsection{Results}
\textbf{The use of transferable knowledge boosts the performance of \ourTool to 74.8\%, a significant improvement over using raw issue-patch pairs (68.3\%). } Upgrading to transferable knowledge from the raw issue-patch pairs improve the performance of \ourTool by 9.7\% (i.e., 68.30\%  \textrightarrow{} 74.80\%). In contrast, using raw issue-patch pairs in \ourTool ($\ourTool_{\text{woK}}$) results in only a slight improvement, from 67.74\% to 68.30\% ($\ourTool_{\text{woT}}$). This remarkable difference underscores the central role of our high-level, transferable knowledge in effectively helping issue resolution.

\textbf{The efficacy of analysis scaling is critically dependent on knowledge guidance and only provides a major performance improvement when guided by knowledge.}
\Cref{tab:ablation} confirms that knowledge guidance and analysis scaling are deeply synergistic. The full system suffers a substantial performance degradation of 3.10\% (74.80\% $\rightarrow$ 72.48\%) when scaling is disabled, underscoring the importance of multi-perspective analysis informed by diverse knowledge. Conversely, without knowledge, enabling analysis scaling yields a statistically negligible improvement of only 0.31\% (67.74\% $\rightarrow$ 67.43\%). This indicates that the utility of analysis scaling is contingent upon being directed by high-quality, transferable knowledge.

In summary, the ablation study demonstrates that transferable knowledge forms the foundation of our framework, and analysis scaling, when combined with it, significantly enhances performance beyond what either component achieves alone.

\subsection{RQ3: Understanding the contribution of Transferable Knowledge within \ourTool}

\subsubsection{Approach}

\begin{figure}
    \centering
    \includegraphics[width=0.8\linewidth]{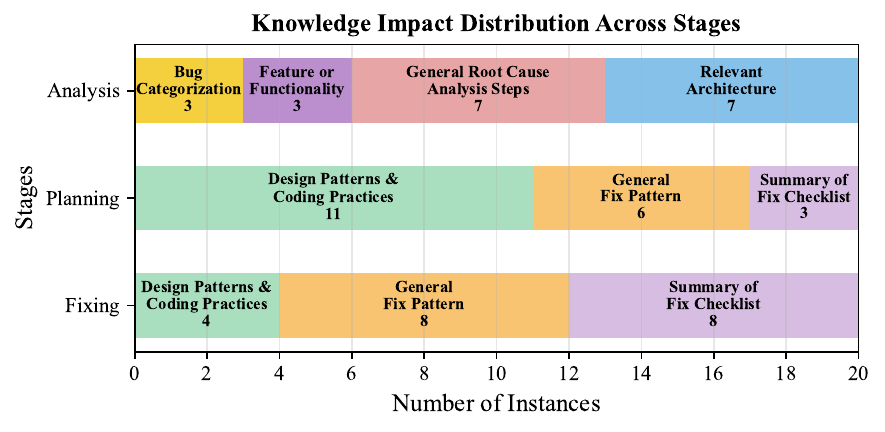}
    \caption{The Contribution of Transferable Procedural Knowledge within the Lingxi}
    \label{fig:rq3}
    \vspace{-0.3cm}
\end{figure}

% In RQ1 and RQ2, our experimental results show that most knowledge-enhanced patches generated by \ourTool achieve reasonable performance.
To further examine the effectiveness of the constructed transferable procedural knowledge and which types of information make the most significant contribution, we compare two resolution trajectories (the logs produced by agents including various information, such as prompts, output, trace of tool calls) of the same sampled SWE-bench instance, produced by $\ourTool$ and $\ourTool_{woK}$ (without transferable knowledge), respectively.

%To isolate knowledge effects, we compare $\ourTool$ against $\ourTool_{woK}$.
%
We selected 20 instances, where \ourTool successfully resolved the issue with the procedural knowledge, while $\ourTool_{woK}$ failed.
Two authors, each with over five years of software development experience, independently analyzed these 20 instances (two trajectories for each, 40 in total).
The analysis focused on the contributions of the knowledge in each stage in the pipeline (i.e., issue analysis, issue fixing planning, and issue fixing) within \ourTool, compared to the one without it. More specifically, the authors analyzed the trajectories and judged which types of information (twelve types of information presented in \Cref{sec:knowledgeConstruction}) in the transferable knowledge make the most significant contribution to issue resolution at each stage.
After completing the reviews, the two authors met to discuss and resolve the discrepancies. The reviewing process achieves a substantial agreement with Cohen's kappa~\cite{cohen1960coefficient} equals 0.66.
%s
%

\begin{table}[t]
\caption{The contributions of Knowledge Aspects on \sweid{sympy-23413}}
\label{tab:knowledge_contrib}
\scriptsize
\centering
\newcommand{\shortindent}{\hspace*{2em}}
\begin{tabularx}{\textwidth}{@{}p{1.9cm}p{0.45\textwidth}p{0.35\textwidth}@{}}
\toprule
\textbf{Task} & \textbf{Key Steps (with Knowledge)} & \textbf{Key Steps (without Knowledge)} \\
\midrule
Fault location &  
 Located to \texttt{\_hermite\_normal\_form(A)}: \newline
 \shortindent Specifically the loop bound (\texttt{rows = min(m,n)}) \textbf{[K1]} \newline
 Risk of negative col index in pivot placement \textbf{[K2]}  
&  
 Incorrect bug localization to the return statement\\
\midrule
Root cause analysis &  
 Loop overshoots columns \textbf{[K3]} for rectangular matrices \newline
 Missing guard for exhausted columns \textrightarrow{} invalid pivots \textbf{[K2]}  
&  
 Incorrectly associates issue with slicing error instead of loop/guard issue \\
\midrule
Fix plan &  
 Process all rows but stop when cols exhausted [K3] \newline
 Add guard: \texttt{if k < 0: k = 0; break} \textbf{[K2]}  
&  
 To remove slicing instead of focusing on the true root cause. \\
\midrule
Patch generation &  
 \texttt{normalforms.py}: change loop bound \textbf{[K3]} \newline
 Insert termination guard \textbf{[K2]}  
&  
 \texttt{normalforms.py}: removed slicing (`return ...') \\
\midrule
Testing &  
 Update expected results: \newline
  Matrix: \texttt{Matrix(3,0,[])} → \texttt{Matrix([[1],[0],[0]])} \textbf{[K4]} \newline
  DM: \texttt{DM([[],[],[]],ZZ)} → \texttt{DM([[1],[0],[0]],ZZ)} \textbf{[K5]}  
&  
 No test updates; degenerate outputs passed \\
\bottomrule
\end{tabularx}
\end{table}

\begin{table}[t]
\caption{Retrieved issues and Associated Transferable Knowledge for \sweid{sympy-23413}}
\label{tab:source_knowledge}
\centering
\scriptsize
\begin{tabularx}{\textwidth}{@{}lllll@{}} 
\toprule
\textbf{Issue ID} & \textbf{Knowledge Idx} & \textbf{Relevant Knowledge}                 & \textbf{Knowledge Contribution}                                                                                                      \\ \midrule
\multirow{3}{*}{Sympy\#10785}          &    K1     & Respect algorithm contract; avoid shortcuts & Rejected full-matrix return; kept pivot semantics                                           \\ \cline{2-5}
          &    K5    & Process full structure, not shape heuristic & Rectangular $m>n$: process all rows                \\ \cline{2-5}
           &     K3    & Ensure Matrix–DM parity                     & Mirrored test updates across paths                                          \\ \midrule
Sympy\#17130           &    K2     & Add boundary guards; prevent invalid pivots & Stopped when columns exhausted; avoid $k<0$                       \\ \midrule

Sympy\#15872           &   K4    & Tests must reflect correct math             & Updated HNF expectation in tests                                                                  \\ \bottomrule

\end{tabularx}
\end{table}
We present a representative example to illustrate the differences of trajectories between \ourTool and $\ourTool_{woK}$. \sweid{sympy-23413} reports an issue where the Hermite Normal Form (HNF) computation produces incorrect results.
\Cref{tab:knowledge_contrib} compares the behavior of $\ourTool$ and $\ourTool_{woK}$.
Specifically, \Cref{tab:source_knowledge} lists the explicit knowledge extracted from the top three most relevant historical issues for this case.
The two root causes contribute to the problem:
1) the loop bound is incorrectly set at \texttt{ min (m, n)}, where \texttt{m} and \texttt{n} are the matrix dimensions of $m \times n$; and
2) the core loop logic lacks a guard to prevent negative column indices during pivot placement.
Without knowledge guidance, the generated patch only addressed the surface-level symptom and failed to resolve the issue.
With knowledge-guided analysis scaling, \ourTool identified the root cause and produced a correct patch.
The patch processed all rows \texttt{(rows = m)} and added a termination guard \texttt{(if k < 0: k=0; break)} to stop pivot placement when columns are exhausted.
This approach preserved the mathematical semantics of HNF and ensured consistent behavior across both the \texttt{Matrix} and \texttt{DomainMatrix} APIs.
Test cases were also updated to reflect the correct expectations, changing degenerate outputs such as \texttt{Matrix(3,0,[])} to the correct pivoted form \texttt{Matrix([[1],[0],[0]])}.

\subsubsection{Results:}
\textbf{Design patterns $\&$ coding practices emerge as the most critical aspects of knowledge}, as they provide concrete guidance on how patches should be designed and implemented (shown in \Cref{fig:rq3}). In contrast, bug categorization and feature or functionality are less impactful. This is because bug categories often group together issues that require highly diverse fixing strategies, making the category label too coarse to guide concrete solutions. Similarly, functionalities may tend to differ substantially between the target and retrieved issues, which reduces the transferability of such knowledge across issues.

\textbf{The importance of knowledge aspects clearly shifts across stages.} In the analysis phase, root cause analysis and architectural knowledge are the two most dominant aspects. This finding aligns with how human developers typically triage and resolve software issues~\cite{ma2025alibaba}, and resonates with prior work showing that deep root cause understanding~\cite{tang2025agent,chen2025swe,ma2024lingma} and architectural awareness~\cite{ma2025alibaba,yang2025enhancing} are more effective than surface-level fixes.
In the planning and fixing phases, other knowledge aspects introduced by \ourTool, such as general fix patterns, fix checklists, and coding practices, become prominent. Design patterns play a particularly critical role in shaping how patches are planned, while general fix patterns and checklists are most influential in ensuring the correctness and completeness of the final implementation.

\section{Discussion}
%In this section, we analyze the failure cases of \ourTool, the comparison with the TTS-based solution to discuss the future work.

\subsection{Analysis of Patch Generation Failures}
\label{sec:discussion-fail}
Our framework demonstrates superior performance on the widely adopted SWE-bench benchmark. However, 25.4\% of instances remain unresolved, highlighting persistent challenges in repo-level issue resolution. We manually conducted a qualitative study to investigate the causes of failures. Our goal was to develop a taxonomy of failure cases that reflects the sophisticated challenges in agent-based issue resolution and facilitates future research.

A total of 20 randomly sampled failure cases were manually analyzed. Whenever a new failure category emerged that was not represented in the initial taxonomy, labeling was temporarily paused and collaboratively reviewed. These discussions allowed the validation of new categories, refinement of the taxonomy, and re-labeling of affected cases, that ensures consistency. 

\paragraph{Error Categories.} 
In our analysis, we identified five major categories of unsuccessful patch generation:
\begin{itemize}
    \item \textbf{Incomplete Fixes (6/20):} Patches partially addressed the issue. For example, in \sweid{astropy-14365}, the generated patch included \texttt{re.compile()} but lacked the parameter \texttt{re.IGNORECASE}, failing to add the necessary case-insensitive handling. 
    \item \textbf{Incorrect Fault Localization (5/20):} The repair was applied to an incorrect file or function. For instance, in \sweid{matplotlib-20676}, the patch modified \texttt{SpanSelector()} instead of the correct \texttt{new\_axes(self, ax)} function. 
    \item \textbf{Semantic Errors (5/20):} Logically incorrect implementations that compiled but contradicted the intended fix. In \sweid{astropy-13236}, the patch added a warning rather than removing the undesired silent auto-transform. 
    \item \textbf{Overfitting Patches (2/20):} Generated code was tailored to specific tests but failed to generalize. In \sweid{astropy-13033}, an overengineered solution attempted to handle every case instead of applying the simple scalar-or-list distinction used in the ground truth patch. 
    \item \textbf{Side-effect Errors (2/20):} The patch resolved the targeted issue but inadvertently disrupted existing functionality. For example, in \sweid{django-12273}, the patch added logic inside \texttt{\_save\_parents()} and \texttt{\_save\_table()}, altering runtime saving behavior rather than addressing the underlying propagation issue.
\end{itemize}

\paragraph{Underlying Causes.}
Beyond classifying errors, we also examined their underlying knowledge-related causes. We identified two primary sources:
\begin{itemize}
    \item \textbf{Knowledge insufficiency (5/20):} The procedural knowledge was incomplete to effectively guide the agent toward a correct repair. 
    \item \textbf{Uninformed autonomy (15/20):} The agent had access to relevant knowledge but disregarded it, relying instead on its own flawed reasoning. 
\end{itemize}

In 15 of the 20 failed patches, the primary cause was \textbf{uninformed autonomy}: even when informative knowledge was available, the agent failed to incorporate it and instead pursued an incorrect path. For example, in \sweid{astropy-13236}, correct practices for resolving the issue were available (\textit{``Examine type handling and conversion processes''} in the \texttt{<general\_root\_cause\_analysis\_steps>} procedural knowledge). However, the agent ignored the auto-conversion code and produced a semantic error. In other cases, the curated knowledge was simply too sparse to direct the agent toward a correct and generalizable solution. This finding underscores that balancing agent autonomy and knowledge guidance remains a central challenge in agent-based issue repair.

\subsection{Enhancing \ourTool with TTS on patch generation}\label{sec:discussion-tts}
Unlike typical TTS-based methods, which apply TTS to the entire pipeline and generate many candidate patches, \ourTool uses experience-guided exploration in the first analysis phase and combines results from different exploration paths.

Since the randomness introduced by TTS can potentially be useful, we also tested \ourTool with the standard TTS setup.
We fixed the results from the analysis phase and applied TTS to the subsequent issue resolution phase to generate three candidate patches.
Previously techniques that use TTS typically employ a learned patch selector to choose the best candidate~\cite{gao2025trae,aggarwal2025dars,augment}.
However, these selectors are often closed-source or described insufficiently for replication.
As this component is not the focus of our study, we assume an oracle selector that can always identify the correct patch if it exists within the TTS-generated candidates.
Under this ideal assumption, the TTS-augmented version of \ourTool solved 396/500 instances (79.2\%) from SWE-bench Verified, representing a 4.6\% absolute improvement over the standard version.
This result indicates a promising potential for enhancing \ourTool by incorporating knowledge in the patch generation phase.
In the future, we plan to extend our knowledge-based exploration to other parts of the pipeline, such as patch generation.
Our broader goal is to guide LLMs more effectively through the exploration space guided by issue-fixing transferable knowledge.

\subsection{Threats to Validity}
\noindent \textbf{Internal Validity}
Threats to internal validity are associated with bias and errors in the experiment.
One potential threat is the absence of an objective ground truth for RQ3, which rely on manual analysis of qualitative understanding of development knowledge and failure reasons and may introduce human bias. To enhance the reliability of these qualitative findings, two senior engineers reviewed and refined the results, achieving substantial inter-rater agreement (a Cohen's kappa of 0.66 for RQ3) through an iterative refinement process. 

Another threat is the quality of LLM-generated procedural knowledge. Our ablation study demonstrates that models using LLM-generated procedural knowledge (\ourTool) significantly outperform those using raw issue-patch ($\ourTool_{woT}$ ), indicating high knowledge quality. Additionally, we implement a two-step extraction process that systematically captures implicit knowledge and incorporates repository exploration tools to ground the generation process in concrete code context, thereby reducing hallucination risks.

%We employ an iterative refinement process where prompts are adjusted based on evaluation feedback, continuing until inter-rater agreement (measured by Fleiss' kappa) reaches substantial agreement (κ > 0.61) and average quality scores exceed 4.0/5.0 across all criteria. This process required three refinement iterations before convergence criteria were met.

\noindent \textbf{External Validity}
% 我们的数据收集过程依赖于原始漏洞信息，fixed filter在目标函数的变更历史中搜索标注的漏洞修复变更，如果目标经历过一样的漏洞修复变更，则认为目标函数是修复过的。因此，如果标注的漏洞修复变更是不准确的（例如漏洞修复包含多个commit，CVE中只标注了其中部分），或者漏洞修复变更内容就是错误的（首先修复，然后回滚再进行二次修复），那么找到的相似但非也可能不准确。
Threats to external validity are related to the generalizability of \ourTool. \ourTool is evaluated on SWE-bench dataset which only contains Python repositories. However, \ourTool is designed to be language-agnostic. It operates by extracting high-level transferable knowledge, rather than parsing language-specific syntax or semantics. While we hypothesize that our method will transfer to other programming languages and issue types, validating this cross-language and cross-ecosystem applicability remains a key direction for future work.
%FVF, which relies on the accuracy and reliability of the publicly available vulnerability sources that we used for data collection.
%角度：我们和很多sota比较，用了多个开源模型（主要从comprehensive comparisons的角度），且在知名数据集。为了我们会扩展数据集

\section{Conclusion}
In this work, we introduce \ourTool, a novel transferable knowledge-guided framework to mitigate the limitations of ad-hoc exploration in repository-level issue resolution.
\ourTool constructs transferable procedural knowledge through a two-step hierarchical abstraction mechanism, and then leverages the knowledge for analysis scaling. 
Our comprehensive evaluation on the SWE-bench Verified benchmark demonstrates that \ourTool significantly outperforms five state-of-the-art open scaffolding baselines with a Pass@1 of 74.6\%, achieving a substantial improvement. Furthermore, ablation studies underscore the critical role of the transferable knowledge and scaling is critically
dependent on knowledge guidance and only achieves significant performance improvement when guided by transferable knowledge. Our qualitative
study further reveals that design patterns \& coding practices are the most critical knowledge aspect, and the roles of different knowledge aspects switch across different stages (i.e., analysis, planning, and fixing).

\section{Data Availability}
We are undergoing the internal process to publish the code, prompts used in this study.

%\section*{Data Availability}

%\section*{Acknowledgment}
\newpage

\bibliographystyle{ACM-Reference-Format}
\bibliography{references}

%%% -*-BibTeX-*-
%%% Do NOT edit. File created by BibTeX with style
%%% ACM-Reference-Format-Journals [18-Jan-2012].

\begin{thebibliography}{51}

%%% ====================================================================
%%% NOTE TO THE USER: you can override these defaults by providing
%%% customized versions of any of these macros before the \bibliography
%%% command.  Each of them MUST provide its own final punctuation,
%%% except for \shownote{} and \showURL{}.  The latter two
%%% do not use final punctuation, in order to avoid confusing it with
%%% the Web address.
%%%
%%% To suppress output of a particular field, define its macro to expand
%%% to an empty string, or better, \unskip, like this:
%%%
%%% \newcommand{\showURL}[1]{\unskip}   % LaTeX syntax
%%%
%%% \def \showURL #1{\unskip}           % plain TeX syntax
%%%
%%% ====================================================================

\ifx \showCODEN    \undefined \def \showCODEN     #1{\unskip}     \fi
\ifx \showISBNx    \undefined \def \showISBNx     #1{\unskip}     \fi
\ifx \showISBNxiii \undefined \def \showISBNxiii  #1{\unskip}     \fi
\ifx \showISSN     \undefined \def \showISSN      #1{\unskip}     \fi
\ifx \showLCCN     \undefined \def \showLCCN      #1{\unskip}     \fi
\ifx \shownote     \undefined \def \shownote      #1{#1}          \fi
\ifx \showarticletitle \undefined \def \showarticletitle #1{#1}   \fi
\ifx \showURL      \undefined \def \showURL       {\relax}        \fi
% The following commands are used for tagged output and should be
% invisible to TeX
\providecommand\bibfield[2]{#2}
\providecommand\bibinfo[2]{#2}
\providecommand\natexlab[1]{#1}
\providecommand\showeprint[2][]{arXiv:#2}

\bibitem[Aggarwal et~al\mbox{.}(2025)]%
        {aggarwal2025dars}
\bibfield{author}{\bibinfo{person}{Vaibhav Aggarwal}, \bibinfo{person}{Ojasv Kamal}, \bibinfo{person}{Abhinav Japesh}, \bibinfo{person}{Zhijing Jin}, {and} \bibinfo{person}{Bernhard Sch{\"o}lkopf}.} \bibinfo{year}{2025}\natexlab{}.
\newblock \showarticletitle{Dars: Dynamic action re-sampling to enhance coding agent performance by adaptive tree traversal}.
\newblock \bibinfo{journal}{\emph{arXiv preprint arXiv:2503.14269}} (\bibinfo{year}{2025}).
\newblock


\bibitem[Antoniades et~al\mbox{.}(2024)]%
        {antoniades2024swe}
\bibfield{author}{\bibinfo{person}{Antonis Antoniades}, \bibinfo{person}{Albert {\"O}rwall}, \bibinfo{person}{Kexun Zhang}, \bibinfo{person}{Yuxi Xie}, \bibinfo{person}{Anirudh Goyal}, {and} \bibinfo{person}{William Wang}.} \bibinfo{year}{2024}\natexlab{}.
\newblock \showarticletitle{Swe-search: Enhancing software agents with monte carlo tree search and iterative refinement}.
\newblock \bibinfo{journal}{\emph{arXiv preprint arXiv:2410.20285}} (\bibinfo{year}{2024}).
\newblock


\bibitem[{Augment code}(2025)]%
        {augment}
\bibfield{author}{\bibinfo{person}{{Augment code}}.} \bibinfo{year}{2025}\natexlab{}.
\newblock \bibinfo{title}{augment: The most powerful AI software development platform backed by the industry-leading context engine.}
\newblock
\urldef\tempurl%
\url{https://www.augmentcode.com}
\showURL{%
\tempurl}
\newblock
\shownote{accessed 2025-06-10}.


\bibitem[Bertram et~al\mbox{.}(2010)]%
        {bertram2010communication}
\bibfield{author}{\bibinfo{person}{Dane Bertram}, \bibinfo{person}{Amy Voida}, \bibinfo{person}{Saul Greenberg}, {and} \bibinfo{person}{Robert Walker}.} \bibinfo{year}{2010}\natexlab{}.
\newblock \showarticletitle{Communication, collaboration, and bugs: the social nature of issue tracking in small, collocated teams}. In \bibinfo{booktitle}{\emph{Proceedings of the 2010 ACM conference on Computer supported cooperative work}}.
\newblock


\bibitem[Bettenburg et~al\mbox{.}(2008)]%
        {bettenburg2008makes}
\bibfield{author}{\bibinfo{person}{Nicolas Bettenburg}, \bibinfo{person}{Sascha Just}, \bibinfo{person}{Adrian Schr{\"o}ter}, \bibinfo{person}{Cathrin Weiss}, \bibinfo{person}{Rahul Premraj}, {and} \bibinfo{person}{Thomas Zimmermann}.} \bibinfo{year}{2008}\natexlab{}.
\newblock \showarticletitle{What makes a good bug report?}. In \bibinfo{booktitle}{\emph{Proceedings of the 16th ACM SIGSOFT International Symposium on Foundations of software engineering}}.
\newblock


\bibitem[Bissyand{\'e} et~al\mbox{.}(2013)]%
        {bissyande2013got}
\bibfield{author}{\bibinfo{person}{Tegawend{\'e}~F Bissyand{\'e}}, \bibinfo{person}{David Lo}, \bibinfo{person}{Lingxiao Jiang}, \bibinfo{person}{Laurent R{\'e}veillere}, \bibinfo{person}{Jacques Klein}, {and} \bibinfo{person}{Yves Le~Traon}.} \bibinfo{year}{2013}\natexlab{}.
\newblock \showarticletitle{Got issues? who cares about it? a large scale investigation of issue trackers from github}. In \bibinfo{booktitle}{\emph{2013 IEEE 24th international symposium on software reliability engineering (ISSRE)}}. IEEE.
\newblock


\bibitem[Briski(2025)]%
        {test.time.scaling.nvidia}
\bibfield{author}{\bibinfo{person}{Kari Briski}.} \bibinfo{year}{2025}\natexlab{}.
\newblock \bibinfo{title}{{How Scaling Laws Drive Smarter, More Powerful AI}}.
\newblock
\urldef\tempurl%
\url{https://blogs.nvidia.com/blog/ai-scaling-laws/}
\showURL{%
\tempurl}
\newblock
\shownote{accessed 2025-05-25}.


\bibitem[Chaparro et~al\mbox{.}(2017)]%
        {chaparro2017detecting}
\bibfield{author}{\bibinfo{person}{Oscar Chaparro}, \bibinfo{person}{Jing Lu}, \bibinfo{person}{Fiorella Zampetti}, \bibinfo{person}{Laura Moreno}, \bibinfo{person}{Massimiliano Di~Penta}, \bibinfo{person}{Andrian Marcus}, \bibinfo{person}{Gabriele Bavota}, {and} \bibinfo{person}{Vincent Ng}.} \bibinfo{year}{2017}\natexlab{}.
\newblock \showarticletitle{Detecting missing information in bug descriptions}. In \bibinfo{booktitle}{\emph{Proceedings of the 2017 11th joint meeting on foundations of software engineering}}.
\newblock


\bibitem[Chen et~al\mbox{.}(2021)]%
        {chen2021pathidea}
\bibfield{author}{\bibinfo{person}{An~Ran Chen}, \bibinfo{person}{Tse-Hsun Chen}, {and} \bibinfo{person}{Shaowei Wang}.} \bibinfo{year}{2021}\natexlab{}.
\newblock \showarticletitle{Pathidea: Improving information retrieval-based bug localization by re-constructing execution paths using logs}.
\newblock \bibinfo{journal}{\emph{IEEE Transactions on Software Engineering}} (\bibinfo{year}{2021}).
\newblock


\bibitem[Chen et~al\mbox{.}(2025)]%
        {chen2025swe}
\bibfield{author}{\bibinfo{person}{Silin Chen}, \bibinfo{person}{Shaoxin Lin}, \bibinfo{person}{Xiaodong Gu}, \bibinfo{person}{Yuling Shi}, \bibinfo{person}{Heng Lian}, \bibinfo{person}{Longfei Yun}, \bibinfo{person}{Dong Chen}, \bibinfo{person}{Weiguo Sun}, \bibinfo{person}{Lin Cao}, {and} \bibinfo{person}{Qianxiang Wang}.} \bibinfo{year}{2025}\natexlab{}.
\newblock \showarticletitle{SWE-Exp: Experience-Driven Software Issue Resolution}.
\newblock \bibinfo{journal}{\emph{arXiv preprint arXiv:2507.23361}} (\bibinfo{year}{2025}).
\newblock


\bibitem[Cohen(1960)]%
        {cohen1960coefficient}
\bibfield{author}{\bibinfo{person}{Jacob Cohen}.} \bibinfo{year}{1960}\natexlab{}.
\newblock \showarticletitle{A coefficient of agreement for nominal scales}.
\newblock \bibinfo{journal}{\emph{Educational and psychological measurement}} (\bibinfo{year}{1960}).
\newblock


\bibitem[Ding et~al\mbox{.}(2014)]%
        {ding2014knowledge}
\bibfield{author}{\bibinfo{person}{Wei Ding}, \bibinfo{person}{Peng Liang}, \bibinfo{person}{Antony Tang}, {and} \bibinfo{person}{Hans Van~Vliet}.} \bibinfo{year}{2014}\natexlab{}.
\newblock \showarticletitle{Knowledge-based approaches in software documentation: A systematic literature review}.
\newblock \bibinfo{journal}{\emph{Information and Software Technology}} (\bibinfo{year}{2014}).
\newblock


\bibitem[Gallant(2015)]%
        {ripgrep}
\bibfield{author}{\bibinfo{person}{Andrew Gallant}.} \bibinfo{year}{2015}\natexlab{}.
\newblock \bibinfo{title}{{GitHub REST API documentation}}.
\newblock
\urldef\tempurl%
\url{https://github.com/BurntSushi/ripgrep}
\showURL{%
\tempurl}
\newblock
\shownote{accessed 2025-09-05}.


\bibitem[Gao et~al\mbox{.}(2024)]%
        {gao2024llm}
\bibfield{author}{\bibinfo{person}{Jingtong Gao}, \bibinfo{person}{Bo Chen}, \bibinfo{person}{Xiangyu Zhao}, \bibinfo{person}{Weiwen Liu}, \bibinfo{person}{Xiangyang Li}, \bibinfo{person}{Yichao Wang}, \bibinfo{person}{Zijian Zhang}, \bibinfo{person}{Wanyu Wang}, \bibinfo{person}{Yuyang Ye}, \bibinfo{person}{Shanru Lin}, {et~al\mbox{.}}} \bibinfo{year}{2024}\natexlab{}.
\newblock \showarticletitle{Llm-enhanced reranking in recommender systems}.
\newblock \bibinfo{journal}{\emph{arXiv preprint arXiv:2406.12433}} (\bibinfo{year}{2024}).
\newblock


\bibitem[Gao et~al\mbox{.}(2025)]%
        {gao2025trae}
\bibfield{author}{\bibinfo{person}{Pengfei Gao}, \bibinfo{person}{Zhao Tian}, \bibinfo{person}{Xiangxin Meng}, \bibinfo{person}{Xinchen Wang}, \bibinfo{person}{Ruida Hu}, \bibinfo{person}{Yuanan Xiao}, \bibinfo{person}{Yizhou Liu}, \bibinfo{person}{Zhao Zhang}, \bibinfo{person}{Junjie Chen}, \bibinfo{person}{Cuiyun Gao}, {et~al\mbox{.}}} \bibinfo{year}{2025}\natexlab{}.
\newblock \showarticletitle{Trae Agent: An LLM-based Agent for Software Engineering with Test-time Scaling}.
\newblock \bibinfo{journal}{\emph{arXiv preprint arXiv:2507.23370}} (\bibinfo{year}{2025}).
\newblock


\bibitem[Ge et~al\mbox{.}(2023)]%
        {ge2023openagi}
\bibfield{author}{\bibinfo{person}{Yingqiang Ge}, \bibinfo{person}{Wenyue Hua}, \bibinfo{person}{Kai Mei}, \bibinfo{person}{Juntao Tan}, \bibinfo{person}{Shuyuan Xu}, \bibinfo{person}{Zelong Li}, \bibinfo{person}{Yongfeng Zhang}, {et~al\mbox{.}}} \bibinfo{year}{2023}\natexlab{}.
\newblock \showarticletitle{Openagi: When llm meets domain experts}.
\newblock \bibinfo{journal}{\emph{Advances in Neural Information Processing Systems}} (\bibinfo{year}{2023}).
\newblock


\bibitem[Hindle et~al\mbox{.}(2008)]%
        {hindle2008large}
\bibfield{author}{\bibinfo{person}{Abram Hindle}, \bibinfo{person}{Daniel~M German}, {and} \bibinfo{person}{Ric Holt}.} \bibinfo{year}{2008}\natexlab{}.
\newblock \showarticletitle{What do large commits tell us? a taxonomical study of large commits}. In \bibinfo{booktitle}{\emph{Proceedings of the 2008 international working conference on Mining software repositories}}.
\newblock


\bibitem[Jimenez et~al\mbox{.}(2023)]%
        {jimenez2023swe}
\bibfield{author}{\bibinfo{person}{Carlos~E Jimenez}, \bibinfo{person}{John Yang}, \bibinfo{person}{Alexander Wettig}, \bibinfo{person}{Shunyu Yao}, \bibinfo{person}{Kexin Pei}, \bibinfo{person}{Ofir Press}, {and} \bibinfo{person}{Karthik Narasimhan}.} \bibinfo{year}{2023}\natexlab{}.
\newblock \showarticletitle{Swe-bench: Can language models resolve real-world github issues?}
\newblock \bibinfo{journal}{\emph{arXiv preprint arXiv:2310.06770}} (\bibinfo{year}{2023}).
\newblock


\bibitem[Kalliamvakou et~al\mbox{.}(2014)]%
        {kalliamvakou2014promises}
\bibfield{author}{\bibinfo{person}{Eirini Kalliamvakou}, \bibinfo{person}{Georgios Gousios}, \bibinfo{person}{Kelly Blincoe}, \bibinfo{person}{Leif Singer}, \bibinfo{person}{Daniel~M German}, {and} \bibinfo{person}{Daniela Damian}.} \bibinfo{year}{2014}\natexlab{}.
\newblock \showarticletitle{The promises and perils of mining github}. In \bibinfo{booktitle}{\emph{Proceedings of the 11th working conference on mining software repositories}}.
\newblock


\bibitem[Ko et~al\mbox{.}(2006)]%
        {ko2006exploratory}
\bibfield{author}{\bibinfo{person}{Amy~J Ko}, \bibinfo{person}{Brad~A Myers}, \bibinfo{person}{Michael~J Coblenz}, {and} \bibinfo{person}{Htet~Htet Aung}.} \bibinfo{year}{2006}\natexlab{}.
\newblock \showarticletitle{An exploratory study of how developers seek, relate, and collect relevant information during software maintenance tasks}.
\newblock \bibinfo{journal}{\emph{IEEE Transactions on software engineering}} (\bibinfo{year}{2006}).
\newblock


\bibitem[Koziolek et~al\mbox{.}(2024)]%
        {koziolek2024llm}
\bibfield{author}{\bibinfo{person}{Heiko Koziolek}, \bibinfo{person}{Sten Gr{\"u}ner}, \bibinfo{person}{Rhaban Hark}, \bibinfo{person}{Virendra Ashiwal}, \bibinfo{person}{Sofia Linsbauer}, {and} \bibinfo{person}{Nafise Eskandani}.} \bibinfo{year}{2024}\natexlab{}.
\newblock \showarticletitle{LLM-based and retrieval-augmented control code generation}. In \bibinfo{booktitle}{\emph{Proceedings of the 1st International Workshop on Large Language Models for Code}}.
\newblock


\bibitem[Li et~al\mbox{.}(2024)]%
        {li2024long}
\bibfield{author}{\bibinfo{person}{Tianle Li}, \bibinfo{person}{Ge Zhang}, \bibinfo{person}{Quy~Duc Do}, \bibinfo{person}{Xiang Yue}, {and} \bibinfo{person}{Wenhu Chen}.} \bibinfo{year}{2024}\natexlab{}.
\newblock \showarticletitle{Long-context llms struggle with long in-context learning}.
\newblock \bibinfo{journal}{\emph{arXiv preprint arXiv:2404.02060}} (\bibinfo{year}{2024}).
\newblock


\bibitem[Liu et~al\mbox{.}(2023)]%
        {liu2023lost}
\bibfield{author}{\bibinfo{person}{Nelson~F Liu}, \bibinfo{person}{Kevin Lin}, \bibinfo{person}{John Hewitt}, \bibinfo{person}{Ashwin Paranjape}, \bibinfo{person}{Michele Bevilacqua}, \bibinfo{person}{Fabio Petroni}, {and} \bibinfo{person}{Percy Liang}.} \bibinfo{year}{2023}\natexlab{}.
\newblock \showarticletitle{Lost in the middle: How language models use long contexts}.
\newblock \bibinfo{journal}{\emph{arXiv preprint arXiv:2307.03172}} (\bibinfo{year}{2023}).
\newblock


\bibitem[Lu et~al\mbox{.}(2022)]%
        {lu2022reacc}
\bibfield{author}{\bibinfo{person}{Shuai Lu}, \bibinfo{person}{Nan Duan}, \bibinfo{person}{Hojae Han}, \bibinfo{person}{Daya Guo}, \bibinfo{person}{Seung-won Hwang}, {and} \bibinfo{person}{Alexey Svyatkovskiy}.} \bibinfo{year}{2022}\natexlab{}.
\newblock \showarticletitle{Reacc: A retrieval-augmented code completion framework}.
\newblock \bibinfo{journal}{\emph{arXiv preprint arXiv:2203.07722}} (\bibinfo{year}{2022}).
\newblock


\bibitem[Ma et~al\mbox{.}(2024)]%
        {ma2024lingma}
\bibfield{author}{\bibinfo{person}{Yingwei Ma}, \bibinfo{person}{Rongyu Cao}, \bibinfo{person}{Yongchang Cao}, \bibinfo{person}{Yue Zhang}, \bibinfo{person}{Jue Chen}, \bibinfo{person}{Yibo Liu}, \bibinfo{person}{Yuchen Liu}, \bibinfo{person}{Binhua Li}, \bibinfo{person}{Fei Huang}, {and} \bibinfo{person}{Yongbin Li}.} \bibinfo{year}{2024}\natexlab{}.
\newblock \showarticletitle{Lingma swe-gpt: An open development-process-centric language model for automated software improvement}.
\newblock \bibinfo{journal}{\emph{arXiv preprint arXiv:2411.00622}} (\bibinfo{year}{2024}).
\newblock


\bibitem[Ma et~al\mbox{.}(2025)]%
        {ma2025alibaba}
\bibfield{author}{\bibinfo{person}{Yingwei Ma}, \bibinfo{person}{Qingping Yang}, \bibinfo{person}{Rongyu Cao}, \bibinfo{person}{Binhua Li}, \bibinfo{person}{Fei Huang}, {and} \bibinfo{person}{Yongbin Li}.} \bibinfo{year}{2025}\natexlab{}.
\newblock \showarticletitle{Alibaba lingmaagent: Improving automated issue resolution via comprehensive repository exploration}. In \bibinfo{booktitle}{\emph{Proceedings of the 33rd ACM International Conference on the Foundations of Software Engineering}}.
\newblock


\bibitem[OpenAI(2024)]%
        {swe.bench.verified}
\bibfield{author}{\bibinfo{person}{OpenAI}.} \bibinfo{year}{2024}\natexlab{}.
\newblock \bibinfo{title}{{Introducing SWE-bench Verified}}.
\newblock
\urldef\tempurl%
\url{https://openai.com/index/introducing-swe-bench-verified/}
\showURL{%
\tempurl}
\newblock
\shownote{accessed 2025-07-10}.


\bibitem[{\"O}rwall(2025)]%
        {moatless}
\bibfield{author}{\bibinfo{person}{Albert {\"O}rwall}.} \bibinfo{year}{2025}\natexlab{}.
\newblock \bibinfo{title}{{moatless-tools}}.
\newblock
\urldef\tempurl%
\url{https://github.com/aorwall/moatless-tools}
\showURL{%
\tempurl}
\newblock
\shownote{accessed 2025-08-25}.


\bibitem[Pabba et~al\mbox{.}(2025)]%
        {pabba2025semagent}
\bibfield{author}{\bibinfo{person}{Anvith Pabba}, \bibinfo{person}{Alex Mathai}, \bibinfo{person}{Anindya Chakraborty}, {and} \bibinfo{person}{Baishakhi Ray}.} \bibinfo{year}{2025}\natexlab{}.
\newblock \showarticletitle{SemAgent: A Semantics Aware Program Repair Agent}.
\newblock \bibinfo{journal}{\emph{arXiv preprint arXiv:2506.16650}} (\bibinfo{year}{2025}).
\newblock


\bibitem[Parvez et~al\mbox{.}(2021)]%
        {parvez2021retrieval}
\bibfield{author}{\bibinfo{person}{Md~Rizwan Parvez}, \bibinfo{person}{Wasi~Uddin Ahmad}, \bibinfo{person}{Saikat Chakraborty}, \bibinfo{person}{Baishakhi Ray}, {and} \bibinfo{person}{Kai-Wei Chang}.} \bibinfo{year}{2021}\natexlab{}.
\newblock \showarticletitle{Retrieval augmented code generation and summarization}.
\newblock \bibinfo{journal}{\emph{arXiv preprint arXiv:2108.11601}} (\bibinfo{year}{2021}).
\newblock


\bibitem[Pl{\"o}sch et~al\mbox{.}(2014)]%
        {plosch2014value}
\bibfield{author}{\bibinfo{person}{Reinhold Pl{\"o}sch}, \bibinfo{person}{Andreas Dautovic}, {and} \bibinfo{person}{Matthias Saft}.} \bibinfo{year}{2014}\natexlab{}.
\newblock \showarticletitle{The value of software documentation quality}. In \bibinfo{booktitle}{\emph{2014 14th International Conference on Quality Software}}. IEEE.
\newblock


\bibitem[Rastkar et~al\mbox{.}(2010)]%
        {rastkar2010summarizing}
\bibfield{author}{\bibinfo{person}{Sarah Rastkar}, \bibinfo{person}{Gail~C Murphy}, {and} \bibinfo{person}{Gabriel Murray}.} \bibinfo{year}{2010}\natexlab{}.
\newblock \showarticletitle{Summarizing software artifacts: a case study of bug reports}. In \bibinfo{booktitle}{\emph{Proceedings of the 32nd ACM/IEEE International Conference on Software Engineering-Volume 1}}.
\newblock


\bibitem[Saff and Ernst(2004)]%
        {saff2004experimental}
\bibfield{author}{\bibinfo{person}{David Saff} {and} \bibinfo{person}{Michael~D Ernst}.} \bibinfo{year}{2004}\natexlab{}.
\newblock \showarticletitle{An experimental evaluation of continuous testing during development}.
\newblock \bibinfo{journal}{\emph{ACM SIGSOFT Software Engineering Notes}} (\bibinfo{year}{2004}).
\newblock


\bibitem[Se and Vert(2025)]%
        {test.time.scaling.huggingface}
\bibfield{author}{\bibinfo{person}{Ksenia Se} {and} \bibinfo{person}{Alyona Vert}.} \bibinfo{year}{2025}\natexlab{}.
\newblock \bibinfo{title}{{What is test-time compute and how to scale it?}}
\newblock
\urldef\tempurl%
\url{https://huggingface.co/blog/Kseniase/testtimecompute}
\showURL{%
\tempurl}
\newblock
\shownote{accessed 2025-05-25}.


\bibitem[Tang et~al\mbox{.}(2025)]%
        {tang2025agent}
\bibfield{author}{\bibinfo{person}{Xiangru Tang}, \bibinfo{person}{Tianrui Qin}, \bibinfo{person}{Tianhao Peng}, \bibinfo{person}{Ziyang Zhou}, \bibinfo{person}{Daniel Shao}, \bibinfo{person}{Tingting Du}, \bibinfo{person}{Xinming Wei}, \bibinfo{person}{Peng Xia}, \bibinfo{person}{Fang Wu}, \bibinfo{person}{He Zhu}, {et~al\mbox{.}}} \bibinfo{year}{2025}\natexlab{}.
\newblock \showarticletitle{Agent kb: Leveraging cross-domain experience for agentic problem solving}.
\newblock \bibinfo{journal}{\emph{arXiv preprint arXiv:2507.06229}} (\bibinfo{year}{2025}).
\newblock


\bibitem[Tao et~al\mbox{.}(2025)]%
        {tao2025code}
\bibfield{author}{\bibinfo{person}{Hongyuan Tao}, \bibinfo{person}{Ying Zhang}, \bibinfo{person}{Zhenhao Tang}, \bibinfo{person}{Hongen Peng}, \bibinfo{person}{Xukun Zhu}, \bibinfo{person}{Bingchang Liu}, \bibinfo{person}{Yingguang Yang}, \bibinfo{person}{Ziyin Zhang}, \bibinfo{person}{Zhaogui Xu}, \bibinfo{person}{Haipeng Zhang}, {et~al\mbox{.}}} \bibinfo{year}{2025}\natexlab{}.
\newblock \showarticletitle{Code Graph Model (CGM): A Graph-Integrated Large Language Model for Repository-Level Software Engineering Tasks}.
\newblock \bibinfo{journal}{\emph{arXiv preprint arXiv:2505.16901}} (\bibinfo{year}{2025}).
\newblock


\bibitem[Wang et~al\mbox{.}(2025)]%
        {wang2025swe}
\bibfield{author}{\bibinfo{person}{Haoran Wang}, \bibinfo{person}{Zhenyu Hou}, \bibinfo{person}{Yao Wei}, \bibinfo{person}{Jie Tang}, {and} \bibinfo{person}{Yuxiao Dong}.} \bibinfo{year}{2025}\natexlab{}.
\newblock \showarticletitle{SWE-Dev: Building Software Engineering Agents with Training and Inference Scaling}.
\newblock \bibinfo{journal}{\emph{arXiv preprint arXiv:2506.07636}} (\bibinfo{year}{2025}).
\newblock


\bibitem[Wang and Lo(2014)]%
        {wang2014version}
\bibfield{author}{\bibinfo{person}{Shaowei Wang} {and} \bibinfo{person}{David Lo}.} \bibinfo{year}{2014}\natexlab{}.
\newblock \showarticletitle{Version history, similar report, and structure: Putting them together for improved bug localization}. In \bibinfo{booktitle}{\emph{Proceedings of the 22nd international conference on program comprehension}}.
\newblock


\bibitem[Wang et~al\mbox{.}(2024)]%
        {wang2024openhands}
\bibfield{author}{\bibinfo{person}{Xingyao Wang}, \bibinfo{person}{Boxuan Li}, \bibinfo{person}{Yufan Song}, \bibinfo{person}{Frank~F Xu}, \bibinfo{person}{Xiangru Tang}, \bibinfo{person}{Mingchen Zhuge}, \bibinfo{person}{Jiayi Pan}, \bibinfo{person}{Yueqi Song}, \bibinfo{person}{Bowen Li}, \bibinfo{person}{Jaskirat Singh}, {et~al\mbox{.}}} \bibinfo{year}{2024}\natexlab{}.
\newblock \showarticletitle{Openhands: An open platform for ai software developers as generalist agents}.
\newblock \bibinfo{journal}{\emph{arXiv preprint arXiv:2407.16741}} (\bibinfo{year}{2024}).
\newblock


\bibitem[Xia et~al\mbox{.}(2024)]%
        {xia2024agentless}
\bibfield{author}{\bibinfo{person}{Chunqiu~Steven Xia}, \bibinfo{person}{Yinlin Deng}, \bibinfo{person}{Soren Dunn}, {and} \bibinfo{person}{Lingming Zhang}.} \bibinfo{year}{2024}\natexlab{}.
\newblock \showarticletitle{Agentless: Demystifying llm-based software engineering agents}.
\newblock \bibinfo{journal}{\emph{arXiv preprint arXiv:2407.01489}} (\bibinfo{year}{2024}).
\newblock


\bibitem[Yang et~al\mbox{.}(2025b)]%
        {yang2025enhancing}
\bibfield{author}{\bibinfo{person}{Boyang Yang}, \bibinfo{person}{Haoye Tian}, \bibinfo{person}{Jiadong Ren}, \bibinfo{person}{Shunfu Jin}, \bibinfo{person}{Yang Liu}, \bibinfo{person}{Feng Liu}, {and} \bibinfo{person}{Bach Le}.} \bibinfo{year}{2025}\natexlab{b}.
\newblock \showarticletitle{Enhancing Repository-Level Software Repair via Repository-Aware Knowledge Graphs}.
\newblock \bibinfo{journal}{\emph{arXiv preprint arXiv:2503.21710}} (\bibinfo{year}{2025}).
\newblock


\bibitem[Yang et~al\mbox{.}(2024)]%
        {yang2024swe}
\bibfield{author}{\bibinfo{person}{John Yang}, \bibinfo{person}{Carlos~E Jimenez}, \bibinfo{person}{Alexander Wettig}, \bibinfo{person}{Kilian Lieret}, \bibinfo{person}{Shunyu Yao}, \bibinfo{person}{Karthik Narasimhan}, {and} \bibinfo{person}{Ofir Press}.} \bibinfo{year}{2024}\natexlab{}.
\newblock \showarticletitle{Swe-agent: Agent-computer interfaces enable automated software engineering}.
\newblock \bibinfo{journal}{\emph{Advances in Neural Information Processing Systems}} (\bibinfo{year}{2024}).
\newblock


\bibitem[Yang et~al\mbox{.}(2025a)]%
        {yang2025swe}
\bibfield{author}{\bibinfo{person}{John Yang}, \bibinfo{person}{Kilian Leret}, \bibinfo{person}{Carlos~E Jimenez}, \bibinfo{person}{Alexander Wettig}, \bibinfo{person}{Kabir Khandpur}, \bibinfo{person}{Yanzhe Zhang}, \bibinfo{person}{Binyuan Hui}, \bibinfo{person}{Ofir Press}, \bibinfo{person}{Ludwig Schmidt}, {and} \bibinfo{person}{Diyi Yang}.} \bibinfo{year}{2025}\natexlab{a}.
\newblock \showarticletitle{Swe-smith: Scaling data for software engineering agents}.
\newblock \bibinfo{journal}{\emph{arXiv preprint arXiv:2504.21798}} (\bibinfo{year}{2025}).
\newblock


\bibitem[Yao et~al\mbox{.}(2023)]%
        {yao2023react}
\bibfield{author}{\bibinfo{person}{Shunyu Yao}, \bibinfo{person}{Jeffrey Zhao}, \bibinfo{person}{Dian Yu}, \bibinfo{person}{Nan Du}, \bibinfo{person}{Izhak Shafran}, \bibinfo{person}{Karthik Narasimhan}, {and} \bibinfo{person}{Yuan Cao}.} \bibinfo{year}{2023}\natexlab{}.
\newblock \showarticletitle{React: Synergizing reasoning and acting in language models}. In \bibinfo{booktitle}{\emph{International Conference on Learning Representations (ICLR)}}.
\newblock


\bibitem[Yin et~al\mbox{.}(2011)]%
        {yin2011fixes}
\bibfield{author}{\bibinfo{person}{Zuoning Yin}, \bibinfo{person}{Ding Yuan}, \bibinfo{person}{Yuanyuan Zhou}, \bibinfo{person}{Shankar Pasupathy}, {and} \bibinfo{person}{Lakshmi Bairavasundaram}.} \bibinfo{year}{2011}\natexlab{}.
\newblock \showarticletitle{How do fixes become bugs?}. In \bibinfo{booktitle}{\emph{Proceedings of the 19th ACM SIGSOFT symposium and the 13th European conference on Foundations of software engineering}}.
\newblock


\bibitem[Zeng et~al\mbox{.}(2025)]%
        {zeng2025satori}
\bibfield{author}{\bibinfo{person}{Guangtao Zeng}, \bibinfo{person}{Maohao Shen}, \bibinfo{person}{Delin Chen}, \bibinfo{person}{Zhenting Qi}, \bibinfo{person}{Subhro Das}, \bibinfo{person}{Dan Gutfreund}, \bibinfo{person}{David Cox}, \bibinfo{person}{Gregory Wornell}, \bibinfo{person}{Wei Lu}, \bibinfo{person}{Zhang-Wei Hong}, {et~al\mbox{.}}} \bibinfo{year}{2025}\natexlab{}.
\newblock \showarticletitle{Satori-SWE: Evolutionary Test-Time Scaling for Sample-Efficient Software Engineering}.
\newblock \bibinfo{journal}{\emph{arXiv preprint arXiv:2505.23604}} (\bibinfo{year}{2025}).
\newblock


\bibitem[Zhang et~al\mbox{.}(2024b)]%
        {zhang2024diversity}
\bibfield{author}{\bibinfo{person}{Kexun Zhang}, \bibinfo{person}{Weiran Yao}, \bibinfo{person}{Zuxin Liu}, \bibinfo{person}{Yihao Feng}, \bibinfo{person}{Zhiwei Liu}, \bibinfo{person}{Rithesh Murthy}, \bibinfo{person}{Tian Lan}, \bibinfo{person}{Lei Li}, \bibinfo{person}{Renze Lou}, \bibinfo{person}{Jiacheng Xu}, {et~al\mbox{.}}} \bibinfo{year}{2024}\natexlab{b}.
\newblock \showarticletitle{Diversity empowers intelligence: Integrating expertise of software engineering agents}.
\newblock \bibinfo{journal}{\emph{arXiv preprint arXiv:2408.07060}} (\bibinfo{year}{2024}).
\newblock


\bibitem[Zhang et~al\mbox{.}(2023)]%
        {zhang2023survey}
\bibfield{author}{\bibinfo{person}{Quanjun Zhang}, \bibinfo{person}{Chunrong Fang}, \bibinfo{person}{Yang Xie}, \bibinfo{person}{Yaxin Zhang}, \bibinfo{person}{Yun Yang}, \bibinfo{person}{Weisong Sun}, \bibinfo{person}{Shengcheng Yu}, {and} \bibinfo{person}{Zhenyu Chen}.} \bibinfo{year}{2023}\natexlab{}.
\newblock \showarticletitle{A survey on large language models for software engineering}.
\newblock \bibinfo{journal}{\emph{arXiv preprint arXiv:2312.15223}} (\bibinfo{year}{2023}).
\newblock


\bibitem[Zhang et~al\mbox{.}(2025b)]%
        {zhang2025and}
\bibfield{author}{\bibinfo{person}{Qiyuan Zhang}, \bibinfo{person}{Fuyuan Lyu}, \bibinfo{person}{Zexu Sun}, \bibinfo{person}{Lei Wang}, \bibinfo{person}{Weixu Zhang}, \bibinfo{person}{Zhihan Guo}, \bibinfo{person}{Yufei Wang}, \bibinfo{person}{Irwin King}, \bibinfo{person}{Xue Liu}, {and} \bibinfo{person}{Chen Ma}.} \bibinfo{year}{2025}\natexlab{b}.
\newblock \showarticletitle{What, how, where, and how well? a survey on test-time scaling in large language models}.
\newblock \bibinfo{journal}{\emph{CoRR}} (\bibinfo{year}{2025}).
\newblock


\bibitem[Zhang et~al\mbox{.}(2025a)]%
        {zhang2025qwen3}
\bibfield{author}{\bibinfo{person}{Yanzhao Zhang}, \bibinfo{person}{Mingxin Li}, \bibinfo{person}{Dingkun Long}, \bibinfo{person}{Xin Zhang}, \bibinfo{person}{Huan Lin}, \bibinfo{person}{Baosong Yang}, \bibinfo{person}{Pengjun Xie}, \bibinfo{person}{An Yang}, \bibinfo{person}{Dayiheng Liu}, \bibinfo{person}{Junyang Lin}, {et~al\mbox{.}}} \bibinfo{year}{2025}\natexlab{a}.
\newblock \showarticletitle{Qwen3 Embedding: Advancing Text Embedding and Reranking Through Foundation Models}.
\newblock \bibinfo{journal}{\emph{arXiv preprint arXiv:2506.05176}} (\bibinfo{year}{2025}).
\newblock


\bibitem[Zhang et~al\mbox{.}(2024a)]%
        {zhang2024autocoderover}
\bibfield{author}{\bibinfo{person}{Yuntong Zhang}, \bibinfo{person}{Haifeng Ruan}, \bibinfo{person}{Zhiyu Fan}, {and} \bibinfo{person}{Abhik Roychoudhury}.} \bibinfo{year}{2024}\natexlab{a}.
\newblock \showarticletitle{Autocoderover: Autonomous program improvement}. In \bibinfo{booktitle}{\emph{Proceedings of the 33rd ACM SIGSOFT International Symposium on Software Testing and Analysis}}.
\newblock


\end{thebibliography}

\end{document}